\documentclass[11pt,letter]{article}
\usepackage[letterpaper, margin=1in,nohead, footskip=36pt]{geometry}
\usepackage[english]{babel}
\usepackage{graphicx}
\usepackage{amsmath}
\usepackage{amsthm}
\usepackage{tabularx,subfigure}
\usepackage[font=footnotesize]{caption}
\usepackage{booktabs}
\usepackage{array}
\usepackage{multirow}
\usepackage{txfonts}
\usepackage{tabularx}
\usepackage{float}
\usepackage{empheq}
\usepackage{xcolor}
\usepackage{footnote}
\usepackage{pstricks, pstricks-add, pst-node, pst-tree, pst-plot, pst-3d}
\usepackage{graphicx, subfigure}

\theoremstyle{plain}
\newtheorem{theorem}{Theorem}[section]
\newtheorem{example}{Example}
\newtheorem{proposition}[theorem]{Proposition}
\newtheorem{corollary}[theorem]{Corollary}
\newtheorem{lemma}[theorem]{Lemma}

\theoremstyle{definition}
\newtheorem{remark}[theorem]{Remark}
\newtheorem{definition}[theorem]{Definition}
\newtheorem{assumption}[theorem]{Assumption}

\floatstyle{ruled}
\newfloat{mip}{h!}{loa}
\floatname{mip}{MIP}

\newfloat{lp}{h!}{loa}
\floatname{lp}{LP}

\newcommand*{\bigtimes}{\mathop{\raisebox{-.5ex}{\hbox{\huge{$\times$}}}}}
\renewcommand{\bigtimes}{\varprod}
\newcommand{\R}{\mathbb{R}}
\newcommand{\N}{\mathbb{N}}
\newcommand{\C}{\mathcal{C}}

\newcommand{\congmodel}{\mathcal{M}}
\newcommand{\Congmodel}{\mathcal{M} = (N,R,X,(c_r)_{r \in R})}

\renewcommand{\d}{\partial}

\newcommand{\pfrac}[2]{#1\,/\,#2}

\newcommand{\rd}{\text{rd}}
\newcommand{\nt}{\text{nt}}
\newcommand{\tr}{\text{t}}
\newcommand{\mc}{\text{MC}}
\newcommand{\twice}{\C^2(\R_{\geq0})}

\newcommand{\no}{[mnode=none]}
\newcommand{\nn}{[mnode=circle, linecolor=white]}

\newcommand{\gr}{[mnode=circle, linecolor=gray]}

\usepackage[draft,marginclue,footnote]{fixme}
\newcommand{\fmth}[2][\empty]{\ifthenelse{\equal{#1}{\empty}}{\fixme{TH:
#2}}{\fixme[#1]{TH: #2}}}
\newcommand{\fmmk}[2][\empty]{\ifthenelse{\equal{#1}{\empty}}{\fixme{MK:
#2}}{\fixme[#1]{MK: #2}}}

\makesavenoteenv{tabular}
\title{~\\
~\\Congestion Games with Variable Demands}

\author{%
  Tobias Harks\thanks{Technische Universit\"at Berlin, Institut
  f\"ur Mathematik, Stra\ss e des 17.~Juni 136, 10623~Berlin,
  Germany. Email: \texttt{\{harks,klimm\}@math.tu-berlin.de}.
  } \and Max Klimm\footnotemark[1]~\thanks{This research was supported
  by the Deutsche Forschungsgemeinschaft within the research training
  group `Methods for Discrete Structures' (GRK 1408).}
}

\date{}

\begin{document}
\maketitle
\begin{abstract}

  We initiate the study of \emph{congestion games with variable
    demands} where the (variable) demand has to be assigned to exactly
  one subset of resources. The players' incentives to use higher
  demands are stimulated by non-decreasing and concave utility
  functions. The payoff for a player is defined as the difference
  between the utility of the demand and the associated cost
  on the used resources. Although this class of non-cooperative
  games captures many elements of real-world applications, it has not
  been studied in this generality, to our knowledge, in the past.

  We study the fundamental problem of the existence of \emph{pure Nash equilibria} (PNE for
  short) in congestion games with variable demands. We call a set
  of cost functions $\C$ \emph{consistent} if every congestion game with variable demands
  and cost functions in $\C$ possesses a PNE. We say that $\C$ is
  \emph{FIP consistent} if every such game possesses the $\alpha$-Finite
  Improvement Property for every $\alpha>0$. Our main results are
  structural characterizations of consistency and FIP consistency for
  twice continuously differentiable cost functions. Specifically, we show
\begin{enumerate}
\item $\C$ is consistent if and only if $\C$ contains either only
affine functions or only homogeneously exponential functions ($c(\ell) = a\,e^{\phi \ell}$).
\item  $\C$ is FIP consistent if and only if $\C$ contains only
affine functions.
\end{enumerate}
Our results provide a  complete characterization of consistency
of cost functions revealing structural differences to
congestion games with \emph{fixed} demands (weighted congestion games),
where in the latter even inhomogeneously exponential functions are
FIP consistent.

Finally, we study consistency and FIP consistency of cost functions in
a slightly different class of games, where every player experiences
the same cost on a resource (uniform cost model). We give a characterization of
consistency and FIP consistency showing that only homogeneously exponential functions
are consistent. 

\iffalse
We obtain the following characterizations:
\begin{enumerate}
\item $\C$ is consistent if and only if $\C$ contains homogeneously exponential functions ($c(\ell) = a\,e^{\phi \ell}$).
\item  $\C$ is FIP consistent consistent if and only if $\C = \emptyset$. \end{enumerate}
\fi

\end{abstract}

\section{Introduction}
Resource allocation problems play a key role in many
applications. Whenever a set of resources needs to be matched to a set
of demands, the goal is to find the most profitable or least costly
allocation of the resources to the demands.  Examples of such
applications come from a wide range of areas, most prominently traffic
networks~\cite{Beckmann56,Roughgarden_Book2005,Smith79,Wardrop52}
and telecommunication networks~\cite{JohariT06,Kelly98,srikant03}.  In
most of the above applications, the allocation of resources is
determined by a finite number of independent players, each optimizing
an individual objective function. A natural framework for analyzing
such non-cooperative games are \emph{congestion games} as introduced
by Rosenthal~\cite{Rosenthal:1973}. In a congestion game, there is a
set of resources and a pure strategy of a player consists of a subset
of resources. The cost of a resource depends only on the number of
players choosing the resource, and the private cost of a player is the
sum of the costs of the chosen resources. Under these 
assumptions, Rosenthal proved the existence of a pure Nash equilibrium
(PNE for short).

In the past, the existence of PNE has been analyzed in many variants
of congestion games such as scheduling, routing, facility location,
network design, each variant with unweighted and weighted players,
see~\cite{Ackermann09,Anshelevich04theprice,Chen06,GairingMT06,Ieong:2005,Milchtaich:2006}.
Most of these previous works have a common feature: given a set of
resources whose cost increases with increasing congestion, every
player allocates a \emph{fixed} demand to an available subset. While
obviously important, such models do not take into account a
fundamental property of many real-world applications: the intrinsic
coupling between the quality or cost of the resources and the
resulting demands for the resources. A prominent example of this
coupling is the flow control problem in telecommunication networks.
In this setting, players receive a non-negative utility from sending
data and the perceived costs increase with congestion.  In this and
other examples, the demands will be reduced if the resources are
congested, and increased if the resources are uncongested. Allowing
for \emph{variable demand} is, thus, a natural prerequisite for
modeling the tradeoff between the benefit for demand, and the quality
of the resource. %(which degrades as the load increases).

There is a large body of work addressing the issue of variable
demands: \cite{Cole06,Low1999,Kelly98,Shenker:1995,srikant03} in the
context of telecommunication networks and~\cite{Beckmann56,Haurie1985}
in the context of traffic networks.  Most of these works assume that
the (variable) demand may be \emph{fractionally} distributed over the
available subsets of resources. This assumption together with
convexity assumptions on the cost and utility functions implies the
existence of a PNE by Rosen's theorem~\cite{Rosen:1965}.
Allowing a fractional distribution of the demand, however, is
obviously not possible in many applications. For instance, the standard TCP/IP protocol suite uses single path routing, because
splitting the demand comes with several practical complications, e.g.,
packets arriving out of order, packet jitter due to different paths delays
etc.

We initiate the study of congestion games with variable demands, where
the (variable) demand has to be assigned to exactly one subset of
resources.  We impose standard economic assumptions (cf.
\cite{Haurie1985,Kelly98,Shenker:1995}) that every player is
associated with a non-decreasing and concave utility function
measuring the utility for the demand. The payoff for a player is
defined as the difference between the utility and the associated
congestion cost on the used resources where the cost of a player on a
resource is the product of the cost function, and the player's demand.

There are two fundamental goals from a system design
perspective: $(i)$ the system must be \emph{stabilizable}, that is,
there must be a stable point (PNE) from which no player wants to
unilaterally deviate; $(ii)$ myopic play of the players should guide
the system to a \emph{stable} state.  
Because the utility functions
are only known to the players (as they are private information) and
not available to the system designer, we study the above two issues
with respect to the used \emph{cost functions} (which represent the
technology associated with the resources, e.g., queuing discipline at
routers, latency function in transportation networks, etc.).  
%More
%formally, we define \emph{PNE-consistency} or simply
%\emph{consistency} of a set of cost functions, see
%also~\cite{Harks:2009c}. 
Let $\mathcal{C}$ be a set of cost functions
and let $\mathcal{G}(\mathcal{C})$ be the set of \emph{all} congestion
games with variable demands and cost functions in $\mathcal{C}$. We
say that $\mathcal{C}$ is \emph{consistent} if every game in
$\mathcal{G}(\mathcal{C})$ possesses a PNE. We say that $\mathcal{C}$
is \emph{FIP consistent} if for every game in $\mathcal{G}(\mathcal{C})$, every sequence of unilateral
$\alpha$-improvements (improvements that add at least $\alpha$ to the
payoff of the deviating player, see~\cite{Monderer:1996}) is finite.
The main goal of this work is to investigate consistency
and the $\alpha$-FIP of the used cost functions.

\paragraph{Connection to Prior Work.}
Already earlier work recognized the importance
of understanding the impact of the \emph{cost structure}
on the existence of PNE and the $\alpha$-FIP
in variants of congestion games, see \cite{Altman02competitiverouting,Fotakis:2005,Libman:2001,OrdaRomSh93,Rosen:1965}. 
The existence of PNE with respect to the cost structure also
plays a fundamental role in the vast literature on Cournot oligopoly games
in the economic theory of imperfect competition.
In a Cournot oligopoly game, there is a set of firms each
producing quantities so as to satisfy an elastic demand. The
production cost for every player is modeled by a convex cost function
and the 
interaction between the firms comes from the price determination mechanism which
is dependent on the total supply on the market.
Note that congestion games with variable demands can be interpreted as a
natural generalization of the Cournot oligopoly game. In~\cite{JT2005} it is proved
that Cournot oligopoly games are basically equivalent (in terms of the
set of Nash equilibria) to congestion games with variable demand and a
single resource (which are termed Cournot oligopsonies
in~\cite{JT2005}). 
The model proposed in this article is more general
since the strategy space of players involves not only quantities 
but also sets of allowable subsets of resources (markets).

The pioneering work of Cournot~\cite{Cournot:1838} established the
existence of an equilibrium point for this model assuming zero costs
and concave inverse demand functions. Further existence results for
more general cost functions have been established by many researchers,
among others~\cite{Novshek:1985,Roberts:1976}. 
 
In the seminal paper by Orda et al.~\cite{OrdaRomSh93},
the authors address the issue of uniqueness of PNE in congestion games with splittable demands. 
They give sufficient conditions for uniqueness of PNE for several classes of cost functions.
In the final section of their paper, the following question is raised (we quote from the paper):
\begin{center}
\emph{
"Several other open questions of practical value deserve attention. For example, in many 
networks users are restricted to route their flow along a single path with strict rules of 
changing them. Under such circumstances an NEP may not exist at all and complicated 
oscillatory behavior is likely to arise."}
\end{center}

\paragraph{Our Results.}

We initiate the study of congestion games with variable demands
and focus on the existence of pure Nash equilibria and the finite improvement
property.
Our main results are structural characterizations of the existence of PNE and the $\alpha$-FIP 
with respect to the cost structure.  Specifically, we show
the following:
\begin{enumerate}
\item 

Let $\C$ be a set of non-negative, strictly increasing and twice
differentiable cost functions. We prove that $\mathcal{C}$ is consistent if and
only if exactly one of the following cases hold: (a)~$\mathcal{C}$
contains only affine functions $c(\ell) = a_c \,\ell + b_c$ with $a_c >0$, $b_c \geq 0$; 
$(b)$ $\C$ contains only homogeneously exponential functions such that $c(\ell) = a_c\,
e^{\phi \ell} $ for some $a_c, \phi>0$, where $a_c$ 
may depend on $c$, while $\phi$ must be equal for all $c \in \C$.
Moreover, we characterize the $\alpha$-FIP in congestion games with variable demands.
We prove that $\mathcal{C}$ is FIP consistent
if and only if $\mathcal{C}$ contains only affine functions.
The formal results appear as Theorems~\ref{thm:consistent} and
\ref{thm:fip_consistent}.

We show that our results remain valid for games with network structure
and, thus, our characterizations settle the open questions raised by Orda et al.~\cite{OrdaRomSh93}.

\item We then investigate a slightly different class of games that we term
 \emph{uniform} games. They differ from the previously studied games
 in the definition of the players' payoff functions. In uniform games
 the cost for a player on a resource is \emph{not} multiplied with the
 demand of that player. For fixed demands, uniform games
 have been studied by many researchers, e.g., \cite{Fotakis:2005,GairingMT06,Ieong:2005, Milchtaich:2006}.
 Considering uniform cost structures is motivated by a series
 of real-world applications. In large-scale telecommunication
 networks, it is highly desirable to charge every player the
 \emph{same} cost regardless of the actual resource consumption of
 every player, because every resource needs only to communicate a
 single value to the players giving rise to an
 efficient and scalable implementation~\cite{srikant03}. In scheduling
 applications, the cost function is frequently used to model the
 achieved makespan which is (under round-robin processing) equal for
 every job on the same resource.

Our second main result provides a complete characterization
of consistency of a set of cost functions for the uniform cost model.
We prove that
$\mathcal{C}$ is consistent w.r.t. uniform cost games if and
only if $\C$ contains homogeneously exponential functions such that $c(\ell) = a_c\,e^{\phi \ell} $ for some $a_c, \phi >0$, where $a_c$ 
may depend on $c$, while $\phi$ must be equal for all $c \in \C$.
Surprisingly, this characterization reveals that
uniform games need not possess a PNE, even if costs
are affine. 
We also characterize the $\alpha$-FIP in the uniform cost model.
We prove that $\mathcal{C}$ is FIP consistent
if and only if $\mathcal{C}=\emptyset$.
For the case of homogeneously exponential cost functions, however,
we derive an improvement dynamic converging to a PNE, thus, showing that the improvement
graphs of the resulting games are weakly acyclic.
\end{enumerate}

Our results are summarized in Table~\ref{tab:results}. All proofs
missing in this extended abstract as well as some examples are
presented in the Appendix.
\begin{table}[t!]
{\footnotesize
  \caption{Existence of PNE and the $\alpha$-FIP in congestion games
    with variable 
    demands with proportional costs and uniform costs.      Note the fundamental structural difference to weighted congestion
    games (with fixed demands) 
    where in the proportional and uniform cost model
    there is always a PNE and the $\alpha$-FIP for affine and
    inhom. exponential cost functions, see
    \cite{Fotakis:2005,Harks:existence,Harks:2009c,Panagopoulou:2006}.}
\label{tab:results}
\begin{center}
  \begin{tabular*}{0.7\textwidth}{@{}l@{\extracolsep{\fill}}c@{\extracolsep{3ex}}c@{\extracolsep{3ex}}c@{\extracolsep{3ex}}c@{}c@{\extracolsep{3ex}}c@{}}
\toprule 
\multicolumn{1}{r}{\multirow{2}{*}{}}
  & \multicolumn{2}{c}{variable demands} & \multicolumn{2}{c}{variable
  demands} & \multicolumn{2}{c}{fixed demands}  \\ 
& \multicolumn{2}{c}{proportional costs} & \multicolumn{2}{c}{uniform
  costs} & \multicolumn{2}{c}{prop. \& unif. costs} \\
  \multicolumn{1}{l}{cost functions } & \multicolumn{1}{c}{PNE} &
  \multicolumn{1}{c}{$\alpha$-FIP} & \multicolumn{1}{c}{PNE} &
  \multicolumn{1}{c}{$\alpha$-FIP} & \multicolumn{1}{c}{PNE} &
  \multicolumn{1}{c}{$\alpha$-FIP} \\ 
  \midrule affine& yes
& yes & no
 & no
 & yes~\cite{Fotakis:2005} &
  yes~\cite{Fotakis:2005}\\
  [1ex] hom. exp. & yes
  & no &
  yes & no
  &
  yes~\cite{Panagopoulou:2006} &
  yes~\cite{Panagopoulou:2006}\\
  \midrule inhom. exp. & no &
  no & no
 & no
  & yes~\cite{Harks:2009c} &
  yes~\cite{Harks:2009c}
  \\[1ex] non aff. \& non exp. & no
 & no
  & no
  & no
  & no~\cite{Harks:existence} &
  no~\cite{Harks:existence} \\
 \bottomrule 
\end{tabular*}
  \end{center}
}
 \end{table}

 \paragraph{Main Ideas and Outline.}
After introducing the basic model in Section~\ref{sec:prel}, 
we prove the "only if" direction of our first result (Theorem~\ref{thm:consistent}) 
in Section~\ref{sec:necessary}. In the proof, we first establish a
connection between weighted congestion games 
and congestion game with variable demands. Given a weighted congestion without PNE, we derive a
congestion game with variable demands using the same cost functions
that also has no PNE. The proof idea relies on a careful
design of feasible (concave and differentiable) utility functions which preserve the improvement
cycles of the original weighted congestion game.  Thus, we can use an
earlier result of~\cite{Harks:existence} stating that a set of cost
functions is consistent for weighted congestion games if and only if
this set contains either affine functions or certain exponential
functions. The hard part of completing the "only if" direction
lies in excluding inhomogeneously exponential cost functions.
We prove that these functions are not consistent by studying
a class of congestion games with fixed \emph{resource dependent demands}.
We identify a subclass of these games with inhomogeneously exponential cost functions
for which we construct a congestion game with variable demands. We show that the thus constructed game
fulfills the invariant of preserved improvement cycles with respect to the original game. While this part of the proof is perhaps the most technical, we obtain as a side-product
of our analysis
a novel characterization of the existence of PNE for congestion games with resource dependent demands
showing that only affine functions are consistent. 

In Section~\ref{sec:sufficient}, the "if" part is proven. We
 introduce a novel potential function concept that we term
 \emph{essential generalized ordinal potentials}. The idea is to require that there is a real-valued function that must increase 
 only for a subset of improving moves. We further introduce
 \emph{local} essential potentials, where this property must 
 hold only for a global maximum of the potential.
 For games with homogeneously exponential
 cost functions, we derive a \emph{local essential potential} completing the "if" direction
 (games with affine costs are exact potential games).
In Section~\ref{sec:uniform}, we investigate a class of games that we term \emph{uniform congestion games with variable demands}. We give similar characterizations
of consistency of cost functions, yielding that only
homogeneously exponential
 cost functions are consistent. We prove the "if" direction by deriving an essential
 potential.  
 We conclude the paper in Section~\ref{sec:conclusions} by presenting new research directions.
 
 \iffalse
 \paragraph{Paper Organization.}
The rest of this paper is organized as follows. 
After introducing the basic model, we present in
Section~\ref{sec:necessary}  the proof of the  ``only if'' parts of 
Theorems~\ref{thm:consistent} and \ref{thm:fip_consistent}. 
%First, we give a structural connection between congestion games with variable demands and %weighted congestion games and derive that any set of consistent or $\alpha$-FIP consistent %cost functions may only contain affine or exponential functions. Then, we show that ex%potential functions are not $\alpha$-FIP consistent and that inhomogenously exponential %functions are not consistent. As a intermediate step of these results we give a complete %characterizations of those cost functions that guarantee the existence of a PNE in %congestion games with fixed but resource dependent demands. 

In Section~\ref{sec:sufficient}, we show the ``if'' part of Theorems~\ref{thm:consistent} and \ref{thm:fip_consistent}, that is, we show that affine functions give rise to the $\alpha$-FIP and the existence of a PNE. Moreover, we prove that homogeneously exponential functions are consistent. In Section~\ref{sec:uniform}, we introduce and examine a class of games that we term \emph{uniform congestion games with variable demands}. We conclude the paper in Section~\ref{sec:conclusions} by presenting new research directions.
\fi

\section{Preliminaries}\label{sec:prel}
Congestion games with variable demands are strategic games $G =
(N,\bar{X},(\pi_i)_{i\in N})$, where $N = \{1,\dots,n\}$ is the
non-empty and finite set of players, $\bar{X} = \bigtimes_{i \in N}
\bar{X}_i$ is the non-empty set of \emph{states} or
\emph{strategy profiles}, and $\pi_i : \bar{X} \to \N$ is the
\emph{individual payoff} function that specifies the payoff value of
player $i$ for each state $\bar{x} \in \bar{X}$.  We define strategies and payoff
functions using the general notion of a congestion model. A tuple
$\mathcal{M} = (N, R, X, (c_r)_{r \in R})$ is called a
\emph{congestion model} if $N$ is a set of players, $R$ is a finite set of \emph{resources}, and
$X = \bigtimes_{i \in N} X_i$ is the set of
\emph{configurations}. For each player $i \in N$, the set $X_i
\subseteq 2^R$ is a finite collection of subsets of
$R$. Every resource $r \in R$ is endowed with a cost function $c_r :
\R_{\geq 0} \to \R_{\geq 0}$.  In a congestion game with variable demands,
every player $i \in N$ is allowed to choose a configuration $x_i
\in X_i$ and a non-negative demand $d_i \in \R_{\geq 0}$ that she
places on all resources in $x_i$. The incentive to use higher demands
is stimulated by a utility function $U_i : \R_{\geq 0} \to \R_{\geq 0}$ that
measures the benefit that player $i$ receives from choosing a certain
demand. Note that this benefit is independent of the resources chosen
and depends solely on the chosen demand.  

Let $\Congmodel$ be a congestion model and let $(U_i)_{i \in N}$ be a
collection of utility functions. We define a congestion game with
variable demands and \emph{proportional costs} as the game
$G(\congmodel) = (N, \bar{X}, \pi)$, where $\bar{X} = (X,\R_{\geq
0})$, $ \pi = (\pi_i)_{i \in N}$ and
$
\pi_i\bigl(x,d\bigr) = U_i(d_i) - \sum_{r \in x_i} d_i c_r \big(\ell_r\bigl(x,d\bigr) \big),
$
and $\ell_r\bigl((x,d)\bigr) = \sum_{j \in N:r \in x_j} d_j$.  We call
$\ell_r\bigl(x,d\bigr)$ the \emph{load} on resource $r$ under strategy
$(x,d)$.  Note that these games are maximization games.  A
configuration $x \in X$ together with a demand profile $d
\in
\R_{\geq 0}^n$ forms a strategy profile $\bar{x}=(x,d)$.  
Note that proportional costs have been used before
in~\cite{Goemans:2005} for the case of weighted congestion games.

We use standard game theory notation; for a set $S\subseteq N$ we
denote by $-S$ its complement and by $\bar{X}_S = \varprod_{i \in S}
\bar{X}_i$ we denote the set of strategy profiles of players in
$S$. Instead of $\bar{X}_{-\{i\}}$ we will write $\bar{X}_{-i}$, and
with a slight abuse of notation we will sometimes write a strategy
profile as $\bar{x} = (\bar{x}_i, \bar{x}_{-i})$ meaning that
$\bar{x}_i \in \bar{X}_i$ and $\bar{x}_{-i} \in \bar{X}_{-i}$.
  
For a constant $\alpha\geq 0$, a pair
$\bigl(\bar{x},(\bar{y}_i,\bar{x}_{-i})\bigr) \in \bar{X}\times
\bar{X}$ is called an \emph{$\alpha$-improving move} of player~$i$ if
$\pi_i(\bar{x}) + \alpha < \pi_i(\bar{y}_i,\bar{x}_{-i})$.  We denote
by $I^\alpha(i)$ the set of $\alpha$-improving moves of player $i \in
N$, and we set $I^\alpha =
\bigcup_{i \in N} I^\alpha(i)$.  We call a sequence of strategy
profiles $\gamma=(x^0,x^1,\dots)$ an \emph{$\alpha$-improvement path}
if every tuple $(x^{k},x^{k+1})\in I^\alpha$.  A strategy profile
$\bar{x} \in \bar{X}$ is a \emph{pure Nash equilibrium}, or PNE for
short, if $(\bar{x},\bar{y}) \notin I^0$ for all $\bar{y} \in
\bar{X}$.  $G$ has the $\alpha$-finite improvement property
($\alpha$-FIP) if every $\alpha$-improvement path is finite.  Let $\C$
be a class of cost functions. We call $\C$ \emph{consistent w.r.t.
  congestion games with variable demands} (or simply
\emph{consistent}) if every congestion game with variable demands and
cost functions in $\C$ admits a PNE. $\C$ is \emph{FIP consistent}
if every congestion game with variable demands and cost functions in
$\C$ has the $\alpha$-FIP.

The following two  assumptions contain mild restrictions on feasible
utility functions and cost functions and are standard in the literature,
see~\cite{Haurie1985,Kelly98,Shenker:1995}.
\begin{assumption}
\label{assumption:cost}
For every resource $r \in R$ the cost function $c_r :
\R_{\geq 0} \to \mathbb{R}_{\geq 0}$ is twice continuously differentiable and strictly increasing. 
\end{assumption}
We denote by $ \twice$ the set of functions satisfying Assumption~\ref{assumption:cost}.
\begin{assumption}
\label{assumption:utility}
 For every player $i \in N$ the utility function $U_i : \R_{\geq 0}
 \to \R_{\geq 0}$ is differentiable, non-decreasing and concave.
\end{assumption}
\begin{remark}In contrast to most of the works in the area of Cournot games or congestion games with splittable
demands (e.g.,\cite{Haurie1985,Kelly98,OrdaRomSh93}), we do not assume 
semi-convexity of cost functions.
\end{remark}

\section{Necessary Conditions}
\label{sec:necessary}
At first we present some simple but useful observations.
It is easy to see that consistent cost functions cannot have bounded marginal costs. Formally, if $\C$ is consistent, then $c(x) + xc'(x) \rightarrow \infty$ as $x \rightarrow \infty$ for all $c \in \C$. To see this, assume that there is $c \in \C$ with $c(x) + xc'(x) < M$ for some $M \in \R_{>0}$. Consider a game with one resource and cost function $c$ and one player with utility function $U(x) = (M+1)\,x$. The payoff of the player increases with the demand, thus, this game does not admit a PNE.

Moreover, the following lemma will be useful
throughout this paper. It simply uses the first-order optimality condition
of a PNE.

\begin{lemma}
\label{lemma:equilibrium_demand}
In any PNE $(x^*,d^*)$ of a congestion game with variable demands $\smash{U_i'(d_i^*) = \frac{\d \,d_i^* \!\sum_{r \in
x_i} c_r(\ell_r(x^*,d^*))}{\d d_i^*}}$ for all $i \in N$ with
$d_i^* > 0$, and $U_i'(d_i^*)\leq \frac{\d \,d_i^* \!\sum_{r \in
x_i} c_r(\ell_r(x^*,d^*))}{\d d_i^*}$ for all $i \in N$ with
$d_i^* =0$.
\end{lemma}

\paragraph{Necessity of Affine or Exponential Costs.}
\label{sec:necessary_affine_exp}
First, we establish a connection between congestion
games with variable demands and  weighted congestion games.
We will exploit this connection to show that consistent
cost functions for congestion games with variable demands must be either affine or exponential.

Let $\mathcal{M} = (N, R, X, (c_r)_{r \in R})$ be a congestion model
and $\bigl(d_i^\text{w}\bigr)_{i \in N}$ be a vector of demands with
$d_i^\text{w} \in \mathbb{R}_{>0}$.  The corresponding \emph{weighted
  congestion game} is the strategic game
$G^\text{w}(\mathcal{M})~=~(N, X, \pi)$, where $\pi$ is defined as
$\pi=\varprod_{i \in N} \pi_i$, $\pi_i(x) = \sum_{r \in x_i}
d_i^\text{w}\,c_r\big(\ell_r(x) \big)$ and $\ell_r(x) = \sum_{j \in N
  : r \in x_j} d_j^\text{w}$.  Note that these games are cost minimization games.
  We will show that whenever there is a weighted congestion game $G^\text{w}(\congmodel)$ that does not admit
a PNE, then there is also a congestion game with variable demands
$G(\congmodel)$ under the same congestion model without a PNE. The proof
of this result is constructive, i.e., given the weighted congestion
game $G^\text{w}(\congmodel)$ and the corresponding vector of demands
$d_i^\text{w}$, we construct a congestion game with variable demands
not possessing a PNE.

Note that the main difficulty lies in the fact that in congestion
games with variable demands the strategy space is strictly larger than
in weighted congestion games. We overcome this issue by designing for
every player $i$ a utility function $U_i$ that allows us to restrict a
priori the set of equilibrium demands of player $i$ to a small
environment $(t_i-\sigma, t_i+\sigma)$ around a \emph{target demand}
$t_i$. To this end, recall that Lemma~\ref{lemma:equilibrium_demand}
establishes that in every PNE player $i$'s marginal utility equals her
marginal cost given that her demand is strictly positive. Our basic
idea is to define the utility function $U_i$ of player~$i$ such that
the following two properties are guaranteed: $(i)$ player~$i$ always has an interest to play a positive demand; $(ii)$ for every
possible value $C'$ of player $i$'s marginal costs, the equation
$U_i'(d_i) = C'$ is met only for $d_i \in (t_i-\sigma, t_i+\sigma)$.
As we will see, we can find smoothed $2$-wise linear utility functions
that suit our purposes.

We define the piecewise linear function through the sequence of points $\bigl((\tau_0,\upsilon_0),(\tau_1,\upsilon_1), \dots, (\tau_k,\upsilon_k)\bigr)$ as the function
\begin{align}
\label{eq:piecewise_linear}
g\bigl((\tau_0,\upsilon_0), \dots, (\tau_k,\upsilon_k)\bigr)\,(x) =
\begin{cases}
\upsilon_{i} + \frac{\upsilon_i-\upsilon_{i-1}}{\tau_i-\tau_{i-1}}(x - \tau_{i}), &\text{ if } x \in [\tau_i, \tau_{i+1}) \text{ for some } i \in \{0,k-1\},\\
\upsilon_k , &\text{ if } x \in [\tau_k,+\infty).
\end{cases}
\end{align}
Since we are interested in obtaining non-decreasing and concave
functions we will assume in the following that the sequence of slopes
$\Bigl(\frac{\upsilon_i-\upsilon_{i-1}}{\tau_i-\tau_{i-1}}\Bigr)_{i=1,\dots,k}$
is positive and non-increasing. To obtain differentiable functions, we
choose a sufficiently small $\sigma>0$ and replace the function $g$ on
every set $(\tau_i-\sigma, \tau_i+\sigma)$, $i=1,\dots,k$ by a
non-decreasing and differentiable function such that the functional
values and first derivatives in $\tau_i-\sigma$ and $\tau_i+\sigma$
comply. For instance, we can choose a suitable quadratic
  function on every interval $(\tau_i-\sigma, \tau_i+\sigma)$, see
  the following example.
  
\begin{example}
  \label{ex:quadratic}
Let $g_{((\tau_1,\upsilon_1), \dots, (\tau_k,\upsilon_k))}$ be a non-decreasing and concave piecewise linear function and let  $0 <\sigma < (\min_{i=1,\dots,k} \tau_i - \tau_{i-1})\,/\,2$ be a smoothening parameter. We regard the function $g^{\sigma}_{((\tau_1,\upsilon_1), \dots, (\tau_k,\upsilon_k))} : \R_{\geq 0} \to \R_{\geq 0}$ defined as
\begin{align*}
g^{\sigma}_{((\tau_1,\upsilon_1), \dots, (\tau_k,\upsilon_k))}\,(x) =
\begin{cases}
\frac{1}{4\sigma}(s_i-s_{i-1})\left(x - \tau_i + \sigma\frac{s_i+s_{i-1}}{s_i-s_{i-1}}\right)^2 \!+\! \upsilon_i \!+\! \sigma s_i \!-\! \frac{\sigma s_i^2}{s_i - s_{i-1}}, &\!\text{if } x \!\in\! (\tau_i \!-\!\sigma, \tau_i \!+\! \sigma), i\!=\!1,\dots,k\\
g_{((\tau_1,\upsilon_1), \dots, (\tau_k,\upsilon_k))}\,(x), &\!\text{else}.
\end{cases}
\end{align*}
Remark that the quadratic functions on $(\tau_i-\sigma,
\tau_i+\sigma)$, $i=1,\dots,k$ are chosen such that the smoothed
piecewise linear function $g^{\sigma}_{((\tau_1,\upsilon_1), \dots,
  (\tau_k,\upsilon_k))}$ is continuously differentiable. Two such functions are shown in Figure~\ref{fig:g}.
  
\begin{figure}[tb]
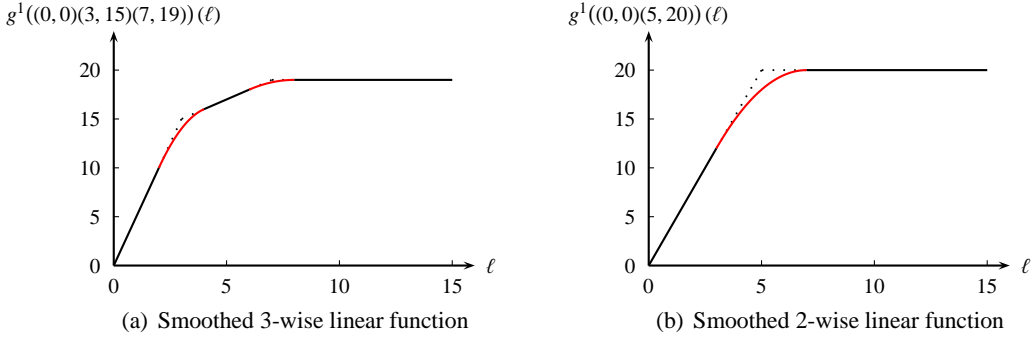

\scriptsize
\hspace{2cm}
\subfigure[Smoothed $3$-wise linear function]{
\psset{xunit=0.3cm,yunit=0.13cm}
\pspicture(-0.5cm,-0.5cm)(18,28)
\psaxes[ticksize=2pt, Dx=5,Dy=5]{->}(0,0)(0,0)(16,24)
\psplot[algebraic=true]{0}{2}{5*x}
\psplot[linestyle=dotted,algebraic=true]{2}{3}{5*x}
\psplot[linestyle=dotted,algebraic=true]{3}{4}{(x-3)+15}
\psplot[linecolor=red, algebraic=true]{2}{4}{1/(4*1)*(1-5)*(x-3+(1*1+1*5)/(1-5))^2 + 3*5 + 1*1-1*1^2/(1-5)}
\psplot[algebraic=true]{4}{6}{(x-4)+16}
\psplot[linestyle=dotted,algebraic=true]{6}{7}{(x-6)+18}
\psplot[linestyle=dotted,algebraic=true]{7}{8}{19}
\psplot[algebraic=true]{8}{15}{19}
\psplot[linecolor=red, algebraic=true]{6}{8}{1/(4*1)*(0-1)*(x-7+(1*0+1*1)/(0-1))^2 + 19 + 1*0-1*0^2/(0-1)}
\rput[l](16.5,0){$\ell$}
\rput[b](0,24.5){$g^{1}\bigl((0,0)(3,15)(7,19)\bigr)\,(\ell)$}
\endpspicture
}
\hspace{1cm}
\subfigure[Smoothed $2$-wise linear function]{
\psset{xunit=0.3cm,yunit=0.13cm}
\pspicture(-0.5cm,-0.5cm)(18,28)
\psaxes[ticksize=2pt, Dx=5,Dy=5]{->}(0,0)(0,0)(16,24)
\psplot{0}{3}{4 x mul}
\psplot[linestyle=dotted]{3}{5}{4 x mul}
\psplot{7}{15}{0 x -5 add mul 20 add}
\psplot[linestyle=dotted]{5}{7}{0 x -5 add mul 20 add}
\psplot[linecolor=red, algebraic=true]{3}{7}{1/(4*2)*(0-4)*(x-5+(2*0+2*4)/(0-4))^2 + 4*5 + 2*0-2*0^2/(0-4)}
\rput[l](16.5,0){$\ell$}
\rput[b](0,24.5){$g^{1}\bigl((0,0)(5,20)\bigr)\,(\ell)$}
\endpspicture
}
\caption{Plot of the smoothed $3$-wise linear function $g^{1}\bigl((0,0)(3,15)(7,19))\bigr)$ and of the smoothed $2$-wise linear function $g^{1}\bigl((0,0)(5,20)\bigr)$.\label{fig:g}}
\end{figure}
\end{example}

We denote the
corresponding smoothed piecewise linear function by
$g^{\sigma}\bigl((\tau_0,\upsilon_0), \dots, (\tau_k,\upsilon_k)\bigr)$. By designing special
smoothed $2$-wise linear functions, we obtain the following
lemma.

\begin{lemma}
\label{thm:target_demand}
Let $\congmodel$ be a congestion model, $(t_i)_{i \in N} \in
\R_{>0}^n$ a vector of target demands with $t_{\min}\!=\!\min_{i \in
  N} \{t_i\}$ and $t_{\max} = \max_{i \in N}\{t_i\}$. Moreover, let $ M >
\max\left\{\frac{2\sum_{r \in R} c_r(T)}{t_{\min}}, \sum_{r \in R}
  \Bigl( c_r(T) + (t_{\max}+\sigma)\max_{\ell \in [t_{\min}, T]}
  c_r'(\ell)\Bigr)\right\}, $ where $T = \sum_{j \in N} (t_j+\sigma)$
and $\sigma < t_{\min}\,/\,2$. Let $G$ be the congestion game with
variable demands and utilities $U_i = g^\sigma_{((0,0),\,(t_i,Mt_i))}$
for all $i \!\in\! N$. Then, $d_i^*\!\in\!(t_i\!-\!\sigma,
t_i\!+\!\sigma)$ for all $i \in N$ and every PNE $(x^*,d^*)$ of $G$.
\end{lemma}

\begin{proof}
  For contradiction, suppose there is a PNE $\bar{x}^* = (x^*, d^*)$
  of $G$ and $i \in N$ with $d_i^* \notin (t_i-\sigma,
  t_i+\sigma)$. First, assume $d_i^* > t_i+\sigma$.  As $c_r$ is strictly increasing, we may assume without loss of generality that there is $r \in x_i$ with $c_r(d_i^*) > 0$. Lemma~\ref{lemma:equilibrium_demand} implies $0=U_i'(d_i^*) =
  \sum_{r \in x_i^*} \bigl(c_r(\ell_r(\bar{x}^*)) +
  d_i^*c'_r(\ell_r(\bar{x}^*)\bigr) >0$, which is a contradiction.

  Now, suppose $d_i^* =0$ and thus $\pi_i(x^*, d^*) =0 $. We denote by $\bar{y} = (x_i^*, t_i-\sigma,
  x_{-i}^*, d_{-i}^*)$ the strategy profile in which player $i$
  chooses her demand equal to $t_i-\sigma$ instead. Note that
  $\pi_i(\bar{y}) = M(t_i - \sigma) - \sum_{r \in x_i^*}
  c_r(\ell_r(\bar{y}))$.  Using that in equilibrium $\smash{d_i^* <
    t_i + \sigma}$, we obtain $\pi_i(\bar{y}) \geq M(t_i - \sigma) -
  \sum_{r \in R} c_r(T),$ which is positive, because $M > 2\sum_{r \in
    R} c_r(T) \,/\, t_{\min} > \sum_{r \in R} c_r(T)\, /\,(t_i -
  \sigma)$. This is a contradiction to the assumption that $d_i^*=0$.

  Finally, let us assume that $0 < d_i^* < t_i-\sigma$. Referring to
  Lemma~\ref{lemma:equilibrium_demand}, we obtain the equality $U_i'(d_i^*) =
  \d\;(d_i^*\sum_{r \in x_i} c_r(\ell(\bar{x}^*))) \, /\, \d d_i^*$.
  In particular,
$$
M = \sum_{r \in x_i} \Bigl(c_r(\ell(\bar{x}^*)) + d_i^*c_r'(\ell(\bar{x}^*))\Bigr) \leq \sum_{r \in R} \Bigl(c_r(T) + (t_i+\sigma)\max_{\ell \in [t_{\min}, T]} c_r'(\ell)\Bigr),
$$
which contradicts $M > \sum_{r \in R} \Bigl( c_r(T) +
(t_i+\sigma)\max_{\ell \in [t_{\min}, T]} c_r'(\ell)\Bigr)$.
\end{proof}

We are now ready to state our first main theorem.
\begin{proposition}\label{pro:weighted-variable}
Let $\congmodel$ be a congestion model and let $G^\text{w} =
(N,X,\pi^\text{w})$ be a corresponding weighted congestion game with
the vector of demands $(d_i^{\text{w}})_{i \in N}$. If $G^\text{w}$
does not admit a PNE, then there exists a congestion game with
variable demands $G(\congmodel)$ to the same congestion model, that
does not admit a PNE.
\end{proposition}

\begin{proof}
  Let $\sigma < \min_{i \in N}(d_i^\text{w})$, $M>1$ and let
  $G(\congmodel)^{M,\sigma} = G^{M,\sigma}$ be a parameterized
  congestion game (the parameters are $M,\sigma$) with variable demands
  and utility functions $\smash{U_i^{M,\sigma} =
    g^\sigma\bigl((0,0),\,(d_i^\text{w},M d_i^\text{w})\bigr)}$. We claim
  that for sufficiently small $\sigma$ and sufficiently large $M$,
  there is a game $G(\congmodel)^{M,\sigma}$ that does not admit a
  PNE.

  First note that the set $X$ of strategy profiles in $G^\text{w}$
  equals the set of configuration profiles in $G^{M,\sigma}$. Since
  $G^\text{w}$ does not admit a PNE, we can find for every strategy
  profile $x \in X$ a player $i(x) \in N$ and $y(x) \in X$ such that
  $\pi_{i(x)}^{\text{w}}(x) > \pi_{i(x)}^\text{w}(y(x))$ and $y(x) =
  (y_i(x),x_{-i})$. Note that weighted congestion games are
  minimization games and, thus, player~$i(x)$ improves her cost when
  switching from strategy $x_i$ to $y_i(x)$. Because the set $X$ of
  strategy profiles of $G^\text{w}$ is finite, we obtain $\delta =
  \min_{ x\in X} \bigl(\pi_{i(x)}(x) - \pi_{i(x)}(y(x))\bigr) > 0$.

  For a
  contradiction, assume that there is a PNE $(x^*, d^*)$ of
  $G^{M,\sigma}$. Assuming that $M$ is sufficiently large and
  referring to Lemma~\ref{thm:target_demand}, we obtain $d_i^* \in
  (d_i^\text{w} - \sigma, d_i^\text{w} + \sigma)$ as a necessary
  condition. Recall that in the weighted congestion game $G^\text{w}$,
  the deviation from $x^*$ to $y(x^*)$ is profitable for player
  $i(x^*)$. Writing $i$ shorthand for $i(x^*)$, we obtain $$
  \pi^\text{w}_{i}\bigl(y(x^*)\bigr) - \pi^\text{w}_i\bigl(x^*\bigr) =
  \sum_{r \in x_i^*} c_r(\ell_r(x^*) - \sum_{r \in y_i(x^*)}
  c_r(\ell_r(y(x^*))= \sum_{r \in x_i^*} c_r\Bigl(\!\!\!\sum_{j \in N
    : r \in x^*_j}\!\!\!d_j^\text{w}\Bigr)- \sum_{r \in y_i(x^*)}
  c_r\Bigl(\!\!\!\sum_{j \in N : r \in y(x^*)_j}\!\!\!
  d_j^\text{w}\Bigr) >\delta.
$$ 
Said differently, the function $h : \R^n_{\geq 0} \to \R$, $d \mapsto
\sum_{r \in x_i^*} c_r\Bigl(\sum\nolimits_{j \in N : r \in x^*_j}
d_j\Bigr)- \sum_{r \in y_i(x^*)} c_r\Bigl(\sum\nolimits_{j \in N : r
  \in y(x^*)_j} d_j\Bigr)$ is positive for $d^\text{w}$.  Note that
$h$ is continuous in every component implying that there is $\epsilon
> 0$ such that $h(d) = \sum_{r \in x_i^*} c_r\Bigl(\sum\nolimits_{j
  \in N : r \in x^*_j} d_j\Bigr)- \sum_{r \in y_i(x^*)}
c_r\Bigl(\sum\nolimits_{j \in N : r \in y(x^*)_j} d_j\Bigr) >0$ for
all $d \in \R^n_{\geq0}$ with $|d_i - d_i^\text{w}| < \epsilon$. We
set $\sigma = \epsilon$ and regard again the game $G^{M,\sigma}$ for
which $\pi_{i}\bigl(y(x^*),d^*\bigr) - \pi_i\bigl(x^*,d^*\bigr) =
\sum_{r \in x_i^*} c_r(\ell_r(x^*,d^*)) - \sum_{r \in y_i(x^*)}
c_r(\ell_r(y(x),d^*) > 0, $ since $d_i^* \in (d_i^\text{w} - \epsilon,
d_i^\text{w} + \epsilon)$. We conclude that player $i$ deviates
profitably from $(x^*,d^*\bigr)$ to $(y(x^*),d^*\bigr)$.  
\end{proof}

In~\cite{Harks:existence} it is shown that for weighted congestion games, a
set $\C$ of (twice continuously differentiable) cost functions is
consistent if and only if one of the following cases holds: $(i)$ $\C$
contains only affine functions; $(ii)$ $\C$ contains only exponential
functions of type $c(\ell) = a_c\, e^{\phi \ell} + b_c$ for some
constants $a_c.b_c,\phi \in \R$ where $a_c$ and $b_c$ may depend on
$c$ while $\phi$ is a common constant for all $c \in \C$.  By
Proposition~\ref{pro:weighted-variable}, we obtain the following result as an
immediate corollary.

\begin{corollary}
\label{cor:hull}
If $\C \subseteq \twice$ is consistent w.r.t.
congestion game with variable demands then one of the following cases
holds: $(i)$ $\C$ contains only affine functions; $(ii)$ $\C$ contains only exponential
functions.
\end{corollary}

The upper result even holds when regarding only network congestion games with variable demands, see the discussion in~\cite{Harks:existence}. Because we only need to consider non-negative and strictly increasing cost functions with unbounded marginal costs, for an affine cost function $c$, we can assume
that $c(\ell) = a\, \ell + b$ with $a > 0$, $b\geq 0$. For
an exponential function  $c(\ell) = a\,e^{\phi \ell} +b$ we can assume $a, \phi>0$ and $b \geq -a$. The function $c$ is called \emph{homogeneously exponential} if $b=0$ and \emph{inhomogeneously exponential} otherwise. Moreover, when we say that a set $\C$ of functions contains only exponential functions, we assume that there is a \emph{universal} constant $\phi>0$ such that every $c \in \C$ can be written as $c(\ell) = a_c\,e^{\phi \ell} + b_c$ for some constants $a_c > 0, b_c \geq -a_c$.

\paragraph{Excluding Inhomogeneously Exponential Costs.}
Next, we show that any inhomogenously exponential function $c(\ell) = a\,e^{\phi \ell} + b$ with $a,\phi >0 $ and $b\neq 0$ is not consistent w.r.t. congestion games with variable demands. In particular, for every such function $c$ there is a congestion game with variable demands and cost equal to $c$ on all resources that does not admit a PNE. In order to prove this result, we first investigate \emph{congestion games with resource dependent demands}. A complete characterization of the set of cost functions that is consistent w.r.t. this class of games is given. We will use a further refinement of this characterization to show that inhomogenously exponential function are not consistent w.r.t. congestion games with variable demands. Let us first define congestion games with resource dependent demands.
\begin{definition}\label{def:resource_dependent}
Let $\mathcal{M} = (N, F, X, (c_r)_{r \in R})$ be a congestion model and let $\left(d_{i,r}\right)_{i \in N,r \in R}$ be a matrix of demands. The corresponding \emph{congestion game with resource dependent demands} is the game $G(\mathcal{M})~=~(N, X, \pi)$, where $\pi$ is defined as $\pi=\varprod_{ i \in N} \pi_i$, $\pi_i(x) = \sum_{r \in x_i}  d_{i,r}\,c_r\big(\ell_r(x) \big)$ and $\ell_r(x) = \sum_{j \in N : r
 \in x_j} d_{j,r}$.
\end{definition}
Restricting the strategy sets to singletons, we obtain scheduling games on unrelated machines as a special case. 
\iffalse
In such game, players are jobs that have
machine-dependent demands and can be scheduled on a set of admissible machines. This problem is often referred to as \emph{restricted scheduling on unrelated machines}. In contrast to the classical approach, where each job strives to minimize its makespan, we consider a different private cost function: Machines charge a price per unit given by a load-dependent cost function $c_r$ and each job minimizes its cost defined as the price of the chosen machine multiplied with its machine-dependent demand. However, our analysis is not restricted to the singleton case.
\fi
Harks et al.~\cite{Harks:2009c} showed that congestion games with resource dependent and affine costs always admit a PNE. Here, we strengthen their result providing a complete characterization of cost functions that are consistent w.r.t. congestion games with resource dependent demand.

\begin{theorem}\label{thm:resource_dependent}
A set $\C \subseteq \twice$ of cost functions is consistent w.r.t. congestion games with resource dependent demands if and only if $\C$ contains only affine functions.
\end{theorem}

\begin{proof}
Since congestion games with resource dependent demands contain weighted congestion games by setting $d_{i,r} = d_{i,r'}$ for all $i \in N$ and $r, r' \in R$, we know that any set of consistent cost functions $\C$ contains either only affine functions or only exponential functions. So it is left to show that for every exponential cost function $c(\ell) = a\, e^{\phi \ell} + b$ with $a,\phi >0$ and $b \geq -a$, there is a congestion game with resource-dependent demands that does not possess a PNE.

To this end, let $c(\ell) = a\,e^{\phi \ell} +b$ with $a,\phi \in \R_{>0}$ and $b \geq -a$ be given and consider the congestion model $\Congmodel$ with three players $N = \{1,2,3\}$ and 28 resources $R = \{r_1,\dots,r_{28}\}$. The strategy spaces of player~1 and 2 contain two strategies only, that is,  $X_1 = \{x_1, \hat{x}_1\}$ and $X_2 = \{x_2,\hat{x}_2\}$. We define $x_1 = \{r_1,\dots,r_8\}$, $\hat{x}_1 = \{r_9,r_{10},r_{11},r_{12}\}$, $x_2 = \{r_{1},\dots,r_{7},r_{9},r_{13},\dots,r_{22}\}$, $\hat{x}_2 = \{r_{8},r_{10},r_{11},r_{23},\dots,r_{28}\}$. Player $3$ has only a single strategy $x_3 = \{r_{8},r_{13},\dots,r_{22}\}$. The strategy spaces are shown in Figure~\ref{fig:resource_dependent}.
 
\setlength{\subfigcapskip}{40pt}
\begin{figure}[tb]
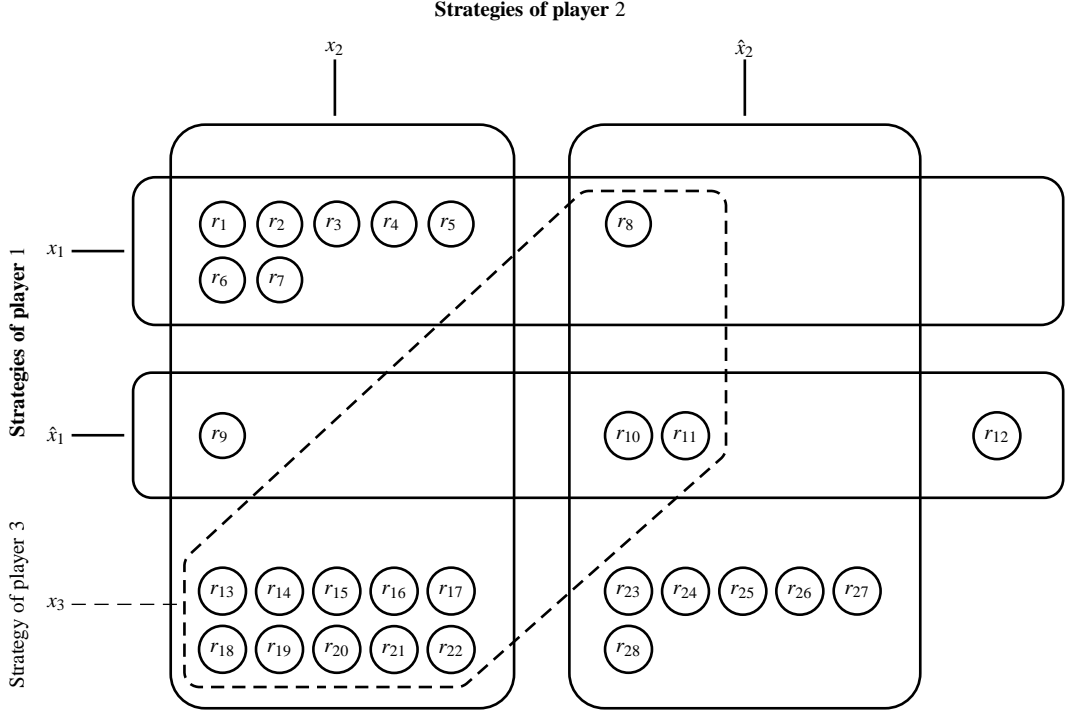

\centering
\vspace{3cm}
% %\subfigure[bla]{
\hspace{0.5cm}
\psset{arrowsize=5pt, arrowlength=1, linewidth=1pt, nodesep=2pt, shortput=tablr}
\scriptsize
\vspace{1cm}
\hspace{2cm}
\begin{psmatrix}[colsep=1mm, rowsep=0.8mm, mnode=circle]
%  1 2 3 4 5 - 1 2 3 4 5 - 1 2 3 4 5
$r_{1~}$&$r_{2~}$&$r_{3~}$&$r_{4~}$&$r_{5~}$&\no\phantom{\hspace{1.5cm}}&$r_{8~}$\\
$r_{6~}$&$r_{7~}$\\
\no\\
\no\\
\no\\
\no\\
$r_{9~}$&        &        &        &        &\no\phantom{\hspace{1.5cm}}&$r_{10}$&$r_{11}$&        &        &       &\no\phantom{\hspace{1cm}}&$r_{12}$\\
\no\\
\no\\
\no\\
\no\\
$r_{13}$&$r_{14}$&$r_{15}$&$r_{16}$&$r_{17}$&\no\phantom{\hspace{1.5cm}}&$r_{23}$&$r_{24}$&$r_{25}$&$r_{26}$&$r_{27}$\\
$r_{18}$&$r_{19}$&$r_{20}$&$r_{21}$&$r_{22}$&\no\phantom{\hspace{1.5cm}}&$r_{28}$\\
\end{psmatrix}
%vertically
\psframe[framearc=0.2](-6.1,-0.8)(-1.4,7.0)
\psframe[framearc=0.2](-11.4,-0.8)(-6.8,7.0)
%horizontally
\psframe[framearc=0.3](-11.9,4.3)(0.5,6.3)	
\psframe[framearc=0.3](-11.9,2.0)(0.5,3.7)
%third player
\pspolygon[linearc=0.2, linestyle=dashed](-11.2,-0.5)(-11.2,1.2)(-5.9,6.1)(-4.0,6.1)(-4.0,2.6)(-7.5,-0.5)
\rput[c]{90}(-13.4,4.1){\bf Strategies of player $1$}
\rput(-12.9,5.3){\bf $x_1$}\psline[linewidth=1pt](-12.7,5.3)(-12.0,5.3)
\rput(-12.9,2.85){\bf $\hat{x}_1$}\psline[linewidth=1pt](-12.7,2.85)(-12.0,2.85)
\rput[c](-6.6,8.5){\bf Strategies of player $2$}
\rput(-9.2,8){\bf $x_2$}\psline[linewidth=1pt](-9.2,7.1)(-9.2,7.85)
\rput(-3.75,8){\bf $\hat{x}_2$}\psline[linewidth=1pt](-3.75,7.1)(-3.75,7.85)
\rput[c]{90}(-13.4,0.6){Strategy of player $3$}
\rput(-12.9,0.6){$x_3$}\psline[linewidth=0.5pt, linestyle=dashed](-12.7,0.6)(-11.3,0.6)
%}
\caption{Strategy spaces of the three players in the game considered in the proof of Theorem~\ref{thm:resource_dependent}. Players $1$ and $2$ have two strategies each, which are drawn as solid boxes. Player $3$ has a single strategy which is shown as a dashed box.\label{fig:resource_dependent}}
\end{figure}

Note that the two strategies of the non-trivial players $i \in \{1,2\}$ are disjoint. The resource-dependent demands of these players are given by $d_{i,r} = \ln 2 \,/\,\phi$ for all resources $r \in x_i$ and $d_{i,r} = 2\ln 2\,/\,\phi$ for all resources $r \in \hat{x}_i$. The demand of player $3$ equals $d_{3,r} = 3\,/\,\phi$ for all $r \in R$.
We calculate that
\begin{align*}
\pi_1(\hat{x}_1,x_2,x_3) - \pi_1(x_1,x_2,x_3) &=  \frac{2 \ln 2}{\phi}\Bigl( 2^2a+b + 2(2^2a+b) + 2^2a+b\Bigl)- \frac{\ln 2}{\phi}\Bigl( 7 (2^2a+b) + 8(2^4a+b\Bigl)\\
&= -124\,a \, \frac{\ln 2}{\phi} < 0.
\end{align*}
Similarly, we obtain 
\begin{align*}
\pi_2(\hat{x}_1, \hat{x}_2,x_3) - \pi_2(\hat{x}_1, x_2,x_3) &= -6\,a\,\frac{\ln 2}{\phi} < 0, & 
\pi_1(x_1, \hat{x}_2,x_3) - \pi(\hat{x}_1,\hat{x}_2,x_3) &= -2\,a\,\frac{\ln 2}{\phi} <0 ,\\
\pi_2(x_1, x_2,x_3) - \pi_2(x_1, \hat{x}_2,x_3) &= -2\,a\,\frac{\ln 2}{\phi}.
\end{align*}
Since the game that we constructed in this proof contains only $4$ strategy profiles and each profile admits a profitable deviation, we conclude that the game does not admit a PNE.
\end{proof}

Let us call a congestion game with resource dependent demands \emph{simple} if each player has a unique demand per strategy, that is, for all $i \in N$ and $x_i \in X_i$ there is $d_{x_i} \in \R_{>0}$ such that $d_{i,r} = d_{x_i}$ for all $r \in x_i$. Since the game constructed in the proof of Theorem~\ref{thm:resource_dependent} is simple, the statement of Theorem~\ref{thm:resource_dependent} is also valid for simple congestion games with resource dependent demands. In particular, for every non-affine function, there is a congestion model $\Congmodel$ and a corresponding simple congestion game with resource dependent demands $G^{\rd} = G^{\rd}(\congmodel)$ with costs equal to $c$ on all resources that has an improvement cycle $(x^1, x^2, \dots, x^s, x^1)$. Now consider a congestion game with variable demands $G = G(\congmodel)$ to the same model, where the utility of each player is constant. Clearly, the cycle $\bigl((x^1,d^1), (x^2,d^2),\dots,(x^s,d^s), (x^1,d^1)\bigr)$, where $d^k_i$ equals the fixed demand of player $i$ in the congestion game with resource dependent demand is also an improvement cycle for $G$. We have established the following.

\begin{proposition}\label{prop:char-alpha-fip}
For a set $\C \subseteq \twice$ the following three are equivalent: $(i)$~$\C$ contains only affine functions; $(ii)$~$\C$ is FIP consistent w.r.t. congestion games with variable demands; $(iii)$~$\C$ is consistent w.r.t. congestion games with resource dependent demands.
\end{proposition}

Note that for any non-affine cost function $c$, there is a congestion game with variable demands and costs equal to $c$ with an improvement cycle. However, there might be PNE outside that cycle.

Our basic idea for  a complete characterization is the following. Let $c$ be an arbitrary exponential function. Using Theorem~\ref{thm:resource_dependent}, there is a simple congestion game with resource dependent demands $G^{rd}$ that does not admit a PNE. We want to construct a congestion game with variable demands $G$ where players' have access on the same resources. We design the players' concave and non-decreasing utility functions such that for any PNE $(x_i, x_{-i}^*,d_i^*, d_{-i}^*)$ in which player $i$ plays configuration $x_i$, the equilibrium demand $d_i^*$ is close to $d_{x_i}$. The key to make this idea work is to introduce additional resources so as to ensure that the players' marginal costs enforce these equilibrium demands. As it turns out, this manipulation can only be done for the case of inhomogenously exponential cost functions. To illustrate this idea, we give an example. If we add two resources with cost $c(\ell) =e^x +1$ to a players' strategy who plays a demand equal to 1, her costs are increased by $2e + 2$ while her marginal costs are increased by $4e+2$. If we instead add one resource with the same cost and a trivial player who plays a demand equal to $\ln(2+e^{-1})$, then her costs are increased by $2e+2$ as well, but her marginal costs are increased by $4e+3$. In this fashion, we can increase the marginal costs of one configurations more than in some other configuration while leaving their differences in costs constant. Since increasing the marginal cost of a configuration, decreases the equilibrium demand, the key challenge is to manipulate the players' marginal costs in order to enforce them play the right equilibrium demands. Before we will develop this idea in the proof of Proposition~\ref{pro:variable_exp}, we need the following technical lemma which is a strengthening of Theorem~\ref{thm:resource_dependent}.

\begin{lemma}
\label{lemma:resource_dependent_delta}
Let $\Delta>0$ be arbitrary. Theorem~\ref{thm:resource_dependent} holds even for simple congestion games with resource dependent demands, where for each player one of the following holds: $(i)$ Player $i$ has exactly one strategy, that is, $X_i = \{x_i\}$ with $x_i \subseteq R$; $(ii)$~Player~$i$ has exactly two strategies that she uses with demands that differ by $\Delta$, that is, $X_i = \{x_i,\hat{x}\}$ and $|d_{x_i} - d_{\hat{x}_i}| = \Delta$ with $x_i, \hat{x}_i \subseteq R$.
\end{lemma}

\begin{proof}
To obtain this more general result, we need a more involved construction compared to that of the proof of Theorem~\ref{thm:resource_dependent}. Let $c(\ell) = a\,e^{\phi \ell} + b$  with $a, \phi \in \R_{>0}$ and $b \geq -a$  and consider for $p,q \in \N_{>0}$ the congestion model $\Congmodel$ with $5$ players and $6q+p$ resources 
\[R = \{r_1, \dots, r_p, s_1, \dots, s_{2q}, v_1, \dots, v_{2q}, w_{1}, \dots, w_{2q}\}.\]
Only the first two players $i \in \{1,2\}$ have two strategies and will be called \emph{non-trivial} players. Their strategy spaces are given by
\begin{align*}
X_1 &= \{x_1, \hat{x}_1\},& &\text{ where } & x_1 &= \{r_1, \dots, r_p, s_1, \dots, s_{2q}\}, & \hat{x}_1 &= \{v_1, \dots, v_{2q}\},\\
X_2 &= \{x_2, \hat{x}_2\},& &\text{ where } & x_2 &= \{r_1, \dots, r_p, w_1, \dots, w_{2q}\}, & \hat{x}_2 &= \{s_1, \dots, s_{2q}\}.
\end{align*}
The players $3$, $4$, and $5$ have only one strategy and serve only to increase the costs on some of the resources. They will be called \emph{trivial players}. Their strategies are given by $X_3 = \bigl\{x_3\bigr\} = \bigl\{\{r_1, \dots, r_p\}\bigr\}$, $X_4 = \bigl\{x_4\bigr\} = \bigl\{s_1, \dots, s_{2q}\}$, $X_5 = \bigl\{x_5\bigr\} = \bigl\{\{v_1, \dots, v_{2q}\}\bigr\}$. The strategy spaces are shown in Figure~\ref{fig:variable_exp} (a). Observe that the strategies of the two non-trivial players are disjoint. For the non-trivial players $i \in \{1,2\}$, we set $d_{i,r} = 2\frac{\ln 2}{\phi}$ for all $r \in x_i$ and $d_{i,r}  = (2 + \frac{p}{q})\frac{ln 2}{\phi}$ for all $r \in \hat{x}_i$.

Assume that the demand of the non-trivial players $3$, $4$, and $5$ are given by $d_{3,r} = \ln(a_3)\,/\,\phi$, $d_{4,r} = \ln(a_4)\,/\,\phi$, and $d_{5,r} = \ln(a_5)\,/\,\phi$ for all $r \in R$, where $a_3, a_4, a_5 >1$ are parameters.  We will carefully choose 
 the parameters $a_3, a_4,$ and $a_5$ such that
\begin{multline*}
\gamma = \bigl((x_1,x_2,x_3,x_4,x_5), (\hat{x}_1,x_2,x_3,x_4,x_5), (\hat{x}_1,\hat{x}_2,x_3,x_4,x_5), (x_1,\hat{x}_2,x_3,x_4,x_5), (x_1,x_2,x_3,x_4,x_5)\bigr)
\end{multline*}
is an improvement cycle. Calculating the costs of the respective deviating non-trivial player in $\gamma$, we obtain the following necessary and sufficient conditions
\begin{align}
2^5\,p\,a_3 + 2^4\,q\,a_4 &> (2+p/q)\,2^{3+p/q}\,q\,a_5 \tag{G1}\label{eq:G1}\\
2^3\,p\, a_3 + 2^4\,q &> (2+p/q)\, 2^{3+p/q}\,q\,a_4 \tag{G2}\label{eq:G2}\\
(2+p/q)\,2^{3+p/q}\,q\,a_5 &> 2^3\,p\,a_3 + 2^{6+p/q}\,q\,a_4 \tag{G3}\label{eq:G3}\\
(2+p/q)\, 2^{5+p/q}\,q\,a_4 &> 2^5\,p\,a_3 + 2^4\,q.\tag{G4} \label{eq:G4}
\end{align}
First, note that the right hand side of inequality \eqref{eq:G1} and the left hand side of inequality \eqref{eq:G3} are equal. This is due to the fact that the second strategy of the first player can be seen as an outside option of player~$1$ since none of the resources in that strategy is contained in one of the strategies of the other non-trivial player~$2$. Thus it is sufficient to consider the inequality
\begin{align*}
2^5\,p\,a_3 + 2^4\,&q\,a_4 > 2^3\,p\,a_3 + 2^{6+p/q}\,q\,a_4\notag\\
\Leftrightarrow\quad &p\,a_3 > \frac{8}{3}\, 2^{p/q}\,q\,a_4 - \frac{2}{3}\,q\,a_4 \tag{G1'}\label{eq:G1prime}
\intertext{instead of \eqref{eq:G1} and \eqref{eq:G3}. Once this conditions is satisfied one can choose $a_5$ appropriately. Inequalities \eqref{eq:G2} and \eqref{eq:G4} give rise to}
p\,a_3 &> 2^{p/q}(2q+p)\,a_4 -2\,q,\tag{G2'}\label{eq:G2prime}\\
 2^{p/q}(2q+p)\,a_4 - q/2 &>\, p\,a_3,\tag{G4'}\label{eq:G4prime}
\end{align*}
respectively. It is left to show that inequalities \eqref{eq:G1prime}, \eqref{eq:G2prime}, and \eqref{eq:G4prime} can be satisfied simultaneously. To this end, observe that the function $f(x) = 2^x(x-\frac{2}{3}) +\frac{2}{3}$ is strictly increasing for $x \geq 0$. As $f(0) = 0$, the function $f$ is strictly positive for every $x>0$. This implies that the inequality $2^{p/q}(\frac{p}{q}-\frac{2}{3}) + \frac{2}{3} > 0$ holds for all $p,q \in \N_{>0}$ 

We define $a_4 > \max\,\{1,\,1\,/\,\bigl(2^{p/q}(\frac{p}{q} - \frac{2}{3}) +\frac{2}{3}\bigr\}$ and $a_3 = 2^{p/q}(2q/p+1)\, a_4\, - \frac{q}{p}$. Since $q>0$, clearly $a_3$ and $a_4$ satisfy \eqref{eq:G2prime} and \eqref{eq:G4prime}. As for \eqref{eq:G1prime}, note that
\begin{align*}
p\, a_3 - \frac{8}{3}\,2^{p/q}\,q\,a_4 + \frac{2}{3}\,q\,a_4 = a_4\,\Biggl(\frac{2}{3}\,q -\frac{2}{3}2^{p/q}\, p - 2^{p/q}\,p\Biggr) - q > 0.
\end{align*}
We set $a_5 = \frac{1}{2(2+p/q)2^{3+p/q}q}(2^5pa_3 + 2^4qa_4 - 2^3pa_3 - 2^{6+p/q}qa_4) = \frac{1}{(2+p/q)2^{3+p/q}q}(12pa_3 + (2^3-2^{5+p/q})qa_4)$. Clearly, $a_5$ satisfies \eqref{eq:G1} and \eqref{eq:G3}. It is left to show, that $a_3,a_4$, and $a_5$ are not smaller than $1$. Since $a_4>1$ by definition, \eqref{eq:G1prime} implies $a_3 > \frac{q}{p}(\frac{8}{3}2^{p/q}-\frac{2}{3}) > 1$. As for $a_5$, inequality \eqref{eq:G3} gives rise to
\begin{align*}
a_5 > \frac{p\,a_3}{(2+\frac{p}{q})\,q\,2^{p/q}} + \frac{8\, a_4}{(2+\frac{p}{q})\,q} = a_4 - \frac{q}{(2+\frac{p}{q})\,q\,2^{p/q}} + \frac{8\, a_4}{(2+\frac{p}{q})\,q} > 1 + \frac{2^{3+p/q}\,q -1}{(2+\frac{p}{q})\, 2^{p/q}} > 1
\end{align*}

Thus, we have found $a_3, a_4, a_5>1$ such that $\gamma$ is an improvement cycle, establishing that the corresponding game does not admit a PNE.
\end{proof}

\setlength{\subfigcapskip}{40pt}
\begin{figure}[p]
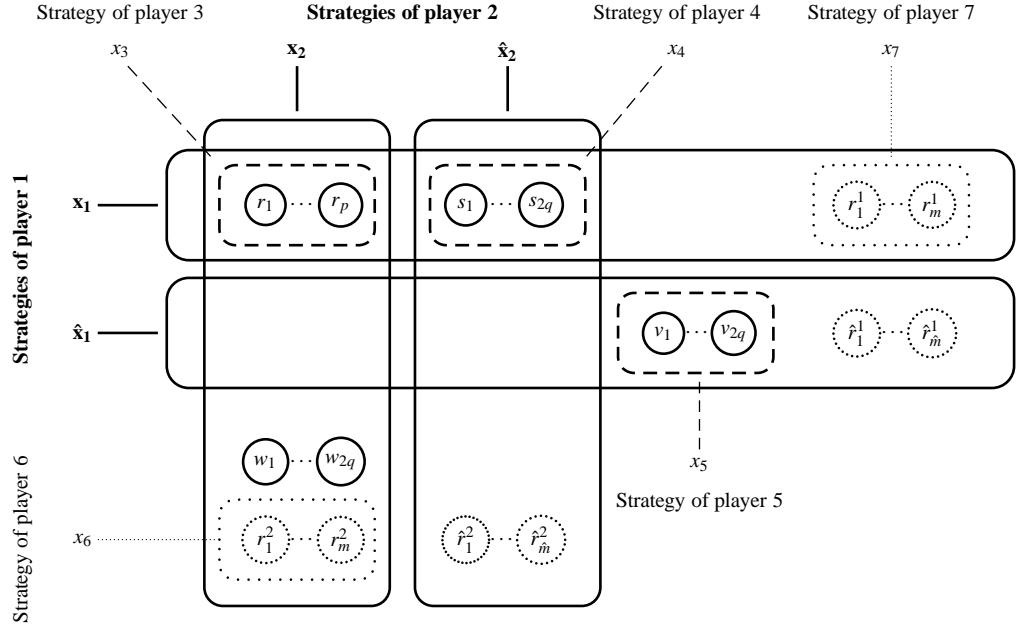
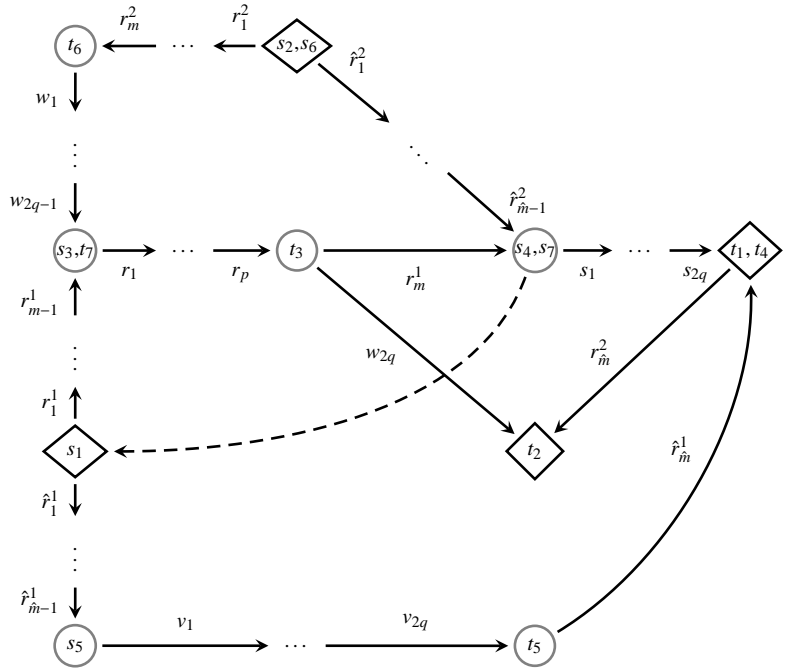

\vspace{2cm}
\subfigure[Non-network congestion game]{
\hspace{4.5cm}
\psset{arrowsize=5pt, arrowlength=1, linewidth=1pt, nodesep=2pt, shortput=tablr}
\scriptsize
\begin{psmatrix}[colsep=0mm, rowsep=0mm, mnode=circle]
%  1 2 3 4 5 - 1 2 3 4 5 - 1 2 3 4 5
$r_1$&\no\dots&$r_p$&\no\phantom{\hspace{1cm}}&$s_1$&\no\dots&$\!s_{2q}\!$&\no\phantom{\hspace{1cm}}& & & &\no\phantom{\hspace{1cm}}&[linestyle=dotted,dotsep=1pt]$\!r_1^1\!$&\no\dots&[linestyle=dotted,dotsep=1pt]$\!r_m^1\!$&\no\\
\no\\
\no\\
\no\\
\no\\
& & &\no\phantom{\hspace{1cm}}& & & &\no\phantom{\hspace{1cm}}&$v_1$&\no\dots&$\!v_{2q}\!$&\no\phantom{\hspace{1cm}}&[linestyle=dotted,dotsep=1pt]$\!\hat{r}_1^1\!$&\no\dots&[linestyle=dotted,dotsep=1pt]$\!\hat{r}_{\hat{m}}^1\!$&\no\\
\no\\
\no\\
\no\\
\no\\
$w_1$&\no\dots&$\!w_{2q}\!$&\no\phantom{\hspace{1cm}}& & & & & & & & & & &\no\\
\no\\
\no\\
[linestyle=dotted,dotsep=1pt]$\!r_1^2\!$&\no\dots&[linestyle=dotted,dotsep=1pt]$\!r_m^2\!$&\no\phantom{\hspace{1cm}}&[linestyle=dotted,dotsep=0.5pt]$\!\hat{r}_1^2\!$&\no\dots&[linestyle=dotted,dotsep=1pt]$\!\hat{r}_{\hat{m}}^2\!$&\no\\
%\no\\
%\no\\
%\no\\
%\no\\
\end{psmatrix}
%vertically
\psframe[framearc=0.2](-7.3,-0.9)(-4.8,5.6)
\psframe[framearc=0.2](-10.1,-0.9)(-7.6,5.6)
%horizontally
\psframe[framearc=0.4](-10.6,3.7)(0.7,5.2)	
\psframe[framearc=0.4](-10.6,2.0)(0.7,3.5)
%player 3
\psframe[linestyle=dashed,framearc=0.4](-9.9,3.9)(-7.8,5.0)
%player 4
\psframe[linestyle=dashed,framearc=0.4](-7.1,3.9)(-5.0,5.0)
%player 5
\psframe[linestyle=dashed,framearc=0.4](-4.6,3.3)(-2.5,2.2)
%player 6
\psframe[linestyle=dotted,framearc=0.4](-9.9,0.55)(-7.8,-0.55)
%player 7
\psframe[linestyle=dotted,framearc=0.4](-2.0,3.9)(0.1,5.0)
%%%
\rput[c]{90}(-12.5,3.6){\bf Strategies of player~1}
\rput(-11.7,4.45){$\bf x_1$}\psline[linewidth=1pt](-11.5,4.45)(-10.7,4.45)
\rput(-11.7,2.75){$\bf \hat{x}_1$}\psline[linewidth=1pt](-11.5,2.75)(-10.7,2.75)
\rput[c](-7.45,7){\bf Strategies of player~2}
\rput(-8.85,6.5){$\bf x_2$}\psline[linewidth=1pt](-8.85,5.7)(-8.85,6.3)
\rput(-6.05,6.5){$\bf \hat{x}_2$}\psline[linewidth=1pt](-6.05,5.7)(-6.05,6.3)
\rput[c](-11.2,7){Strategy of player~3}
\rput(-11.2,6.5){$x_3$}\psline[linewidth=0.5pt,linestyle=dashed](-11.1,6.3)(-10.0,5.1)
\rput[c](-3.8,7){Strategy of player~4}
\rput(-3.8,6.5){$x_4$}\psline[linewidth=0.5pt,linestyle=dashed](-4.0,6.3)(-5.0,5.1)
\rput[c](-3.5,0.5){Strategy of player~5}
\rput(-3.5,1){$x_5$}\psline[linewidth=0.5pt,linestyle=dashed](-3.5,1.2)(-3.5,2.1)
\rput[c]{90}(-12.5,0.0){Strategy of player~6}
\rput(-11.7,0.0){$x_6$}\psline[linewidth=0.5pt, linestyle=dotted, dotsep=1pt](-11.5,0.0)(-9.9,0.0)
\rput[c](-0.95,7){Strategy of player~7}
\rput(-0.95,6.5){$x_7$}\psline[linewidth=0.5pt,linestyle=dotted, dotsep=1pt](-0.95,6.3)(-0.95,5.1)
}
~\\
~\\
~\\
~\\
%new subfigure
\addtolength{\subfigcapskip}{-20pt}
\subfigure[Network congestion game]{
\hspace{3.5cm}
\psset{arrowsize=5pt, arrowlength=1, linewidth=1pt, nodesep=2pt, shortput=tablr}
\scriptsize
\begin{psmatrix}[colsep=7mm, rowsep=5mm, mnode=dia]
%      1                       2                  3                  4         5   6    7        8       9      10        11
\gr$t_6$                  &\nn\dots &$\!\!\!s_2,\!s_6\!\!\!$\\
\nn\vdots                 &         &                          &\nn$\ddots$\\
\gr$\!\!\!s_3,\!t_7\!\!\!$&\nn\dots &\gr$t_3$                  &           &\gr$\!\!\!s_4,\!s_7\!\!\!$ &\nn\dots&$\!\!\!t_1,t_4\!\!\!$\\
\nn\vdots\\
$s_1$ &         &                          &           &$t_2$\\
\nn\vdots \\
\gr$s_5$                     &         &\nn\dots                  &           &\gr$t_5$
\end{psmatrix}
\ncline{->}{1,3}{2,4}^{$\hat{r}_1^2$}
\ncline{->}{1,3}{1,2}^{$r_1^2$}
\ncline{->}{1,2}{1,1}^{$r_{m}^2$}
\ncline{->}{1,1}{2,1}<{$w_1$}
\ncline{->}{2,1}{3,1}<{$w_{2q-1}$}
\ncline{->}{2,4}{3,5}>{$\hat{r}_{\hat{m}-1}^2$}
\ncline{->}{3,1}{3,2}_{$r_1$}
\ncline{->}{3,2}{3,3}_{$r_{p}$}
\ncline{->}{3,3}{3,5}_{$r_m^1$}
\ncline{->}{3,5}{3,6}_{$s_1$}
\ncline{->}{3,6}{3,7}_{$s_{2q}$}
\ncline{->}{3,3}{5,5}<{$w_{2q}$}
\ncline{->}{3,7}{5,5}<{$r_{\hat{m}}^2$}
\ncline{->}{5,1}{4,1}<{$r_1^1$}
\ncline{->}{4,1}{3,1}<{$r_{m-1}^1$}
\ncline{->}{5,1}{6,1}<{$\hat{r}_1^1$}

\ncline{->}{6,1}{7,1}<{$\hat{r}_{\hat{m}-1}^1$}
\ncline{->}{7,1}{7,3}^{$v_1$}
\ncline{->}{7,3}{7,5}^{$v_{2q}$}
\ncarc[arcangle=-30]{->}{7,5}{3,7}<{$\hat{r}_{\hat{m}}^1$}
\ncline{->}{7,5}{9,7}
\ncline{->}{7,9}{9,7}
\nccurve[linestyle=dashed,angleA=-110]{->}{3,5}{5,1}
}

\caption{Strategy spaces of the of the congestion games constructed in the proof of Lemma~\ref{pro:variable_exp} in their non-network and network representations. In the non-network representation (a) the non-trivial players $1$ and $2$ each have two strategies shown by boxes with solid lines while the trivial players $3,4$ and $5$ each have a single strategy drawn as dashed boxes. Omitting the dashed resources as well as the new trivial players $6$ and $7$ one obtains the congestion model constructed in Lemma~\ref{lemma:resource_dependent_delta} on which the proof of Lemma~\ref{pro:variable_exp} is based. In the network representation (b) resources correspond to the edges of the graph. Each player~$i$ is associated with a pair $(s_i,t_i)$ of nodes. The set of strategies of each player equals the set of her directed $(s_i,t_i)$-paths. By adding sufficiently many resources to the path that is indicated by the dashed line, the set of every players undominated non of the non-trivial players will use this path in equilibrium.\label{fig:variable_exp}}
\end{figure} 

We are now ready for the main result of this section.
\begin{proposition}
\label{pro:variable_exp}
Any inhomogenously exponential function is not consistent w.r.t. congestion games with variable demands.
\end{proposition}

\begin{proof}
Let $c(\ell) = a\, e^{\phi \ell} + b$ with $a,\phi >0$ and $b \geq -a$, $b \neq 0$ be an inhomogeneously exponential function. We first assume that $\phi = 1$ and $b>0$. In the concluding remarks of this proof, we sketch how the arguments can be adapted to work with arbitrary $\phi >0$ and $-a \leq b < 0$.

As shown in Lemma~\ref{lemma:resource_dependent_delta}, for every $p,q \in \N_{>0}$ there is a congestion model $\mathcal{M}= (N, R, X, (c_r)_{r \in R})$ with $p + 6q$ resources, $5$ players and cost functions $c_r = c$ for all $r \in R$ such that there is a corresponding congestion game with resource dependent demands $G^\rd = (N,X,\pi^\rd)$ that does not admit a PNE. The game  $G^\rd$ can be constructed to have the following additional properties: The set of players $N^\rd$ is partitioned into two \emph{non-trivial} players $N_{\nt}$ and three \emph{trivial} players $N_{\tr}$. Every trivial player $i \in N_{\tr}$ has a unique strategy $X_i = \{x_i\}$ and a unique demand $t_i = d_{i,r}$ for all $r\in R$. Every non-trivial player $i \in N_{\nt}$ has exactly two disjoint strategies $X_i = \{x_i, \hat{x}_i\}$ and exactly two demands $t_i = 2$ and $\hat{t}_i = 2 + p/q$, such that $d_{i,r} = t_i$ for all $r \in x_i$ and $d_{i,r} = \hat{t}_i$ for all $r \in \hat{x}_i$. For notational convenience, we use the convention $t_i = \hat{t}_i$ for every trivial player $i \in N_{\tr}^\rd$.

Based on the game $G^\rd$ described above, we want to construct a congestion game with variable demands $G$ that does not admit a PNE. Our construction is based on carefully designing the players' utility function. In fact, the players' utilities should offer them an incentive to play (almost) the same demands as in the congestion game with resource dependent demands. That is, we want to assure that in any PNE of $G$ (if it exists) if player~$i$ uses the configuration $x_i$, then her demand is close to $t_i$. In contrast, if player $i$ plays $\hat{x}_i$, then her demand is close to $\hat{t}_i$.

As shown in Lemma~\ref{thm:target_demand}, if $M_i$ is sufficiently large and the utility of the trivial player $i \in N_{\tr}$ is given by the smoothed 2-wise linear function $U_i = g^\sigma\bigl((0,0),\,(t_i,M_it_i)\bigr)$, then $d_i^* \in (t_i-\sigma,t_i+\sigma)$ in every PNE $(x^*,d^*)$ of $G$.

For the non-trivial players, we will use a similar idea. We define their utilities such that $d_i^* \in (t_i-\sigma, t_i + \sigma)$ for every PNE $\bigl((x_i,x_{-i}^*), d^*\bigr)$ in which player $i$ plays her first configuration $x_i$ and $d_i^* \in (\hat{t}_i-\sigma, \hat{t}_i + \sigma)$ for every PNE $\bigl((\hat{x}_i,x_{-i}^*), d^*\bigr)$ in which player $i$ plays her second configuration $\hat{x}_i$. We note the subtle disambiguation between strategies in congestion games (with resources-dependent demands) and strategies in congestion games with variable demands. While a strategy in a congestion game is a subset of resources, a strategy in a congestion game with variable demands is a tuple of a subset of resources and a demand. To distinguish between both, we call the sets $x_i \in X_i \subseteq 2^R$ \emph{strategies} when speaking of the congestion game with resource dependent demand and \emph{configurations} when dealing with the congestion game with variable demand.

To enforce the right demands in the congestion game with variable demands, we will enhance the underlying congestion model. We introduce new resources (parameterized by $m,\hat{m}$) and additional players in Step~2. In Step~3, we will choose the parameters $p,q, m, \hat{m} \in \N_{>0}$ and some $\epsilon>0$. The utility functions of the non-trivial players will be defined in Step~4. Finally, in Step~5 we show that the thus constructed congestion game with variable demand does not admit a PNE.

\paragraph{Step 2: Enhancing the congestion model.} For every non-trivial player $i \in N_{\nt}$ we add $m\in \N$ new resources to $i$'s first strategy $x_i$ and $\hat{m}\in \N$ resources to $i$'s second strategy $\hat{x}_i$. In addition, the $2m$ new resources added to $x_1$ and $x_2$, respectively, shall be used by two new trivial players. Technically, we introduce a new congestion model $\congmodel' = (N', R', X', (c_r)_{r \in R'})$. The new set of players contains two new trivial players. Thus, we define $N'_{\tr} = N_{\tr} \cup \{6,7\}$, $N_{\nt}' = N_{\nt}$, and $N' = N'_{\nt} \cup N'_{\tr}$. 
For every non-trivial player $i \in N'_{\nt}$, we set $x'_i = x_i \cup \{r_1^{i}, \dots, r_m^{i}\}$ and $\hat{x}'_i = \hat{x}_i \cup \{\hat{r}_1^{i}, \dots, \hat{r}_{\hat{m}}^{i}\}$. Note that the new set of resources contains $p+ 6q + 2m + 2\hat{m}$ resources, so
$$R' = \{r_1, \dots, r_p, s_1, \dots, s_{2q}, v_1, \dots, v_{sq}, w_1, \dots, w_{2q}, r_1^1, \dots, r_m^1, \hat{r}_1^1, \dots, \hat{r}_{\hat{m}}^1, r_1^2, \dots, r_m^2, \hat{r}_1^2, \dots, \hat{r}_{\hat{m}}^2\}.$$
The (old) trivial players $3$, $4$, and $5$, have the same strategies as in $G^\rd$. The new trivial player $6$ has a single strategy $x'_6 = \{r_1^1, \dots, r_m^1\}$, the single strategy of player $7$ is $x'_7 = \{r_1^2, \dots, r_m^2\}$. For some $\epsilon>0$ to be defined later, we want these two players to play a demand close to $t_6 = t_7 = p/q + \ln(\frac{3+\epsilon+p/q}{3}) + \ln(\frac{\hat{m}}{m})$ in equilibrium. As we already argued for the other trivial players $3,4$, and $5$ choosing the right smoothed $2$-wise linear function for players $6$ and $7$ their equilibrium demands can be restricted to $d_6^* \in (t_6-\sigma, t_6 + \sigma)$ and $d_7^* \in (t_7 - \sigma, t_7 + \sigma)$, respectively.

Observe that the marginal utility of each player $i \in N'$ is zero for demands $d_i \geq \hat{t}_i + 1/2$. Hence, in any PNE $(x'^*,d^*)$, every player $i \in N'$ will use a demand $d_i^* < \hat{t}_i+1/2$. Let us now consider a non-trivial player $j \in N'_{\nt}$. We observe that $j$'s marginal cost $\mc_j(x_j', x_{-j}^*, t_j, d_{-j}^*) = \d\, t_j \sum_{r \in x_j} c_r\bigl(\ell_r(x_j',x_{-j}^*,t_j, d_{-j}^*)\bigr)\,/\, \d t_j$ when playing her first configuration $x_j'$ and her first target demand $t_j = 2$ can be bounded from below by
\begin{align}
\mc_j(x_j', x'^*_{-j}, t_j, d^*_{-j}) &= 3\sum_{r \in x_j} a\, e^{\ell_r(x_j',x'^*_{-j},t_j, d_{-j}^*)} + \bigl|x_j\bigr|\,b + a\,\hat{m}\, \Biggl(3+\epsilon+\frac{p}{q}\Biggr)\,e^{2+p/q} +m\,b\notag\\
&> a\,\Biggl(3+\frac{p}{q}\Biggr)\,\hat{m}\,e^{2+p/q} + a\,\hat{m}\,\epsilon\,e^{2+p/q} + m\,b. \label{eq:bound1}
\intertext{for all $x'^*_{-j} \in X'_{-j}$ and $d_{-j}^* \in \varprod_{i \neq j} [0,\hat{t}_i+1/2]$. Note that the summation is over $r \in x_i$, that is, we do not sum over the $m$ newly introduced resources. Assuming that no player uses a non-optimal demand (which gives a marginal utility equal to zero) and that $\sigma \leq 1$, player $j$'s marginal costs when playing her first strategy and her first target demand can also be bounded from above by}
\mc_j(x_j', x'^*_{-j}, t_j, d_{-j}^*) &\leq \omega_j + a\,\hat{m}\, \Biggl(3+\epsilon+\frac{p}{q}\Biggr)\,e^{2+p/q} + m\,b, \label{eq:bound2}
\intertext{where $\omega_j = 3\sum_{r \in R} a\, e^{4 + 2p/q + \sum_{j=3,\dots,7} t_j + 7}+ \bigl|R|\,b.$
 Analogously, in every equilibrium $j$'s marginal cost when playing her second configuration $\hat{x}'_j$ and her second target demand $\hat{t}_j = 2+p/q$ can be bounded from above by}
\mc_j(\hat{x}'_j, x'^*_{-j}, \hat{t}_j, d_{-j}^*) &= \Biggl(3+\frac{p}{q}\Biggr)\sum_{r \in \hat{x}_j} a\, e^{\ell_r(\hat{x}_j,x'^*_{-j},\hat{t}_j, d_{-j}^*)} + \bigl|\hat{x}_j\bigr|\,b + \Biggl(3+\frac{p}{q}\Biggr)\,a\,\hat{m}\,e^{2+p/q} + \hat{m}\,b\notag\\
&< \omega_j + a\,\Biggl(3+\frac{p}{q}\Biggr)\,\hat{m}\,e^{2+p/q} + \hat{m}\,b.\label{eq:bound3}
\end{align}
for all $x'^*_{-j} \in X'_{-j}$ and $d^*_{-j} \in \varprod_{i \neq j} [0,\hat{t}_i+1/2]$.

So $j$'s marginal cost when playing her first configuration $x_j'$ with low demand $t_j = 2$ can be bounded from below by $B = a\hat{m}(3+\frac{p}{q})e^{2+p/q} + a\hat{m}\epsilon\,e^{2+p/q}+bm$ and her marginal cost when playing her second configuration $\hat{x}'_j$ with high demand $\hat{t}_j = 2 + p/q$ can be bounded form above by $\hat{B} = \omega_j + a\,\hat{m}\,(3+p/q)e^{2+p/q}+\hat{m}\,b$. Moreover, note that $j$'s actual cost when playing her first configuration $x'_j$ with low demand $t_j$ has been increased by $\frac{2}{3}(3+\epsilon+p/q)\,a\,\hat{m}\,e^{2+p/q} + 2\,b\,m$ by adding $m$ additional resources. In contrast, the cost when playing her second configuration $\hat{x}'_j$ with high demand $\hat{t}_j$ has been increased by $(2+p/q)\,a\,\hat{m}\,e^{2+p/q} + (2+p/q)\,b\hat{m}$. We want that the slope $\tau_j$ of player $j$'s utility between $t_j$ and $\hat{t}_j$ is such that $B > \tau_j > \hat{B}$ in order to give player $j$ the incentive to play $x'_j$ only together with demand $t_j$ and $\hat{x}'_j$ only together with demand $\hat{t}_j$, respectively. 

As we want to cancel out the difference in utility $U_j(\hat{t}_j) - U_j(t_j) \approx \frac{p}{q}\tau_j$ by the additional costs, we require
\begin{align}
\label{eq:tau}
\frac{p}{q}\,\tau_j = \Biggl(2+\frac{p}{q}\Biggr)\,a\,\hat{m}\,e^{2+p/q} + \Biggl(2+\frac{p}{q}\Biggr)\,b\hat{m} - \frac{2}{3}\Biggl(3+\epsilon+\frac{p}{q}\Biggr)\,a\,\hat{m}\,e^{2+p/q} - 2\,b\,m.
\end{align}
 
We proceed showing how to choose $p,q,m,\hat{m}$ and $\epsilon$ in order to satisfy $B>\tau_j>\hat{B}$ and \eqref{eq:tau}.

\paragraph{Step 3: Choosing the parameters $\bf p,q,m,\hat{m} \in \N_{>0}$ and $\bf \epsilon>0$.} Introducing $x = p/q$, we want to solve
\begin{align}
a\,\hat{m}\,(3+x)\,e^{2+x} + a\,\hat{m}\,\epsilon\,e^{2+x} - b(\hat{m}-m)> \Biggl(\frac{1}{3}-\frac{2\epsilon}{3x}\Biggr)\,a\,\hat{m}\,e^{2+x} + \frac{2b}{x}(\hat{m}-m) > a\,\hat{m}\,(3+x)\,e^{2+x} + \omega_i.\label{eq:pqmm}
\end{align}
Defining $\epsilon = \frac{2b}{a}e^{-2-x}$, $m=1$ and dividing by $\hat{m}$ we get
\begin{align}
a\,(3+x)\,e^{2+x} + b + \frac{b}{\hat{m}} > \frac{2}{3}\,b\,x^{-1} + \frac{1}{3}\,a\,\,e^{2+x} - \frac{2b}{x\hat{m}} > a\,(3+x)\,e^{2+x} + \frac{\omega_i}{\hat{m}} \label{eq:final_in}.
\end{align}
The function $f(x) = \frac{2}{3}bx^{-1} + \frac{1}{3}ae^{2+x}-a(3+x)e^{2+x}$ is continuous with respect to $x$. Moreover, $f$ goes to $\infty$ as $x \rightarrow 0$ and goes to $-\infty$ as $x \rightarrow \infty$. In Particular, there is a rational $x = p/q$ such that $b > f(x) > 0$. Hence, $x$ solves \eqref{eq:final_in} for sufficiently large $\hat{m}$.

For the sequel of this proof, let us fix $p,q, m$ and $\hat{m}$ such that they solve \eqref{eq:pqmm}. We continue defining the non-trivial players' utilities.

\paragraph{Step 4: Defining the non-trivial players' utilities.} For every non-trivial player $j \in N'_{\nt}$ we define the utility function $U_j$ as the $3$-wise linear function $U_j = g^\sigma\bigl((0,0),\,(t_j,M_j\,t_j), \, (\hat{t}_j, M_j\,t_j + \tau_j\,(\hat{t}_j - t_j))\bigr)$ with $\tau_j = (\frac{1}{3}-\frac{2\epsilon}{x})\,a\,\hat{m}\,e^{2+x} + \frac{2b}{x}(\hat{m}-m)$ and $M_j > \omega_j + \hat{m}\, (3+\epsilon+p/q)\,e^{2+p/q} + m\,b$. As marginal costs and marginal utilities coincide in any equilibrium, we obtain that $d_j^* \in (t_j-\sigma, t_j + \sigma)$ for every PNE $\bigl((x_j',x_{-j}^*), d^*\bigr)$ in which player $j$ plays her first configuration $x'_j$ and $d_j^* \in (\hat{t}_j-\sigma, \hat{t}_j + \sigma)$ for every PNE $\bigl((\hat{x}_j',x_{-j}^*), d^*\bigr)$ in which player $j$ plays her second configuration $\hat{x}_j'$. We finish the proof showing that we can find a sufficiently small $\sigma$ for which no PNE of $G$ exists.

\paragraph{Step~5: Choosing $\sigma$ such that $G$ does not admit a PNE.}  For contradiction, assume that for every $\sigma>0$ there is a PNE $\bar{x} = ({x'}^*, d^*)$ of $G$. Let us first consider the case $x_1'^* = x_1$ and $x_2'^* = x_2$, that is both non-trivial players use their first strategy. As we argued in the last paragraph, we can deduce $d_i^* \in (t_i - \sigma, t_i + \sigma)$ for all $i \in N'$ as a necessary condition on the equilibrium demand.

We claim that player $1$ improves her utility when switching from $\bar{x} = (x_1', d^*_1)$ to $\bar{y} = (\hat{x}_1',\hat{t}_1)$. Note that
\begin{align*}
\pi_1(\bar{y}) - \pi_1(\bar{x})&= U_1(\hat{t}_1) - \sum_{r \in \hat{x}_1} c_r(\ell_r(\bar{y})) - a\,\hat{m}\,\Biggl(2+\frac{p}{q}\Biggr)\, e^{2+p/q} - b\,\hat{m}\,\Biggl(2+\frac{p}{q}\Biggr)\\
&\quad -\Biggl(U_1(d_1^*) - \sum_{r \in x_1} c_r(\ell_r(\bar{x})   - a\,\hat{m}\,\Biggl(2 + \frac{2\epsilon}{3} + \frac{2p}{3q}\Biggr)\,e^{2+p/q} - 2\,b\,m\Biggr)\\
&= U_1(\hat{t}_1) - U_1(d^*_1) - \Biggl(\frac{p}{3q} - \frac{2\epsilon}{3}\Biggr)\,a\,\hat{m}\,e^{2+p/q} - 2b(\hat{m}-m) - \frac{2p}{q}\,b\hat{m}+ \sum_{r \in x_1} c_r(\ell_r(\bar{x}) - \sum_{r \in \hat{x}_1} c_r(\ell_r(\bar{y})) 
\end{align*}

Since in the congestion game with resource dependent demands $G^\rd$, the deviation from $(x_1, x_2, \dots, x_5)$ to $(\hat{x}_1,x_2,\dots, x_5)$ is profitable for player $1$, there is $\delta >0$ such that the sum $\sum_{r \in x_1} c_r(\ell_r(\bar{x}) - \sum_{r \in \hat{x}_1} c_r(\ell_r(\bar{y})) > \delta >0$. Moreover, note that player $1$'s utility function is chosen such that $U_1(\hat{t}_1) - U_1(d_1^*)$ goes to $\frac{p}{q}\tau = (\frac{p}{3q}-\frac{2\epsilon}{3})\,a\,\hat{m}\,e^{2+x} + 2b(\hat{m}-m) + \frac{2p}{q}b\hat{m}$ as $\sigma$ goes to zero. We conclude that also in the congestion game with variable demands, player $1$ deviates profitable from $\bar{x} = (x_1', d^*_1)$ to $\bar{y} = (\hat{x}_1',\hat{t}_1)$ for $\sigma$ sufficiently small. This contradicts our assumption and we derive that $G$ does not admit a PNE.

\paragraph{Concluding remarks.} With the above construction we have established that for every cost function $c(\ell) = a\,e^{\phi \ell} + b$ with $\phi =1$ and $b>0$, there is a congestion game with variable demands and cost equal to $c$ on all resources that does not possess a PNE.

We sketch the argumentation for $-a \leq b < 0$. The construction is the same except that in Step~3, we require that $\hat{m} < m$ and we set the target demands of the new trivial players to $t_6 = t_7 = p/q + \ln(\frac{3+\epsilon+p/q}{3}) >0$. It is easy to verify that the inequalities \eqref{eq:bound1}, \eqref{eq:bound2}, and \eqref{eq:bound3} stay valid. Then, as $b<0$ and $\hat{m} - m < 0$, we can solve inequality \eqref{eq:pqmm} in Step~3 with the same argument as before. Steps~4 and 5 remain as in the original proof.

For arbitrary $\phi > 0$, we scale the target demands $t_i$ and $\hat{t}_i$ of every player by $1/\phi$.
\end{proof}

\begin{remark}
The above result also holds for network congestion games, see 
the construction in Figure~\ref{fig:variable_exp}~(b).
\end{remark}

Below, we can restrict our search space for consistent cost functions to affine functions and homogeneously exponential functions. In the next section, we will show that both classes of functions are consistent.

\section{Consistent Cost Functions}
\label{sec:sufficient}

\paragraph{Affine Functions.}
We will first show below that games with affine cost functions are exact potential games
and, thus, they possess a PNE and the $\alpha$-FIP for every
$\alpha>0$.

\begin{proposition}
\label{pro:potential} 
Let $G$ be a congestion game with variable demands and affine cost
functions $(c_r)_{r\in R}$.  Then, the function $P(x,d)= \sum_{i \in
N} U_i(d_i) - \sum_{r \in x_i} c_r\left(\sum_{j \in \{1,\dots,i\} : r
\in x_j}d_j \right)d_i$ is an exact potential function.
Thus, $G$ possesses a PNE and the $\alpha$-FIP for
every $\alpha>0$.
\end{proposition}

\begin{proof}
  We consider affine cost functions, that is, $c_r(\ell) = a_r\ell +
  b_r$, $a_r>0, b_r\geq 0$.  We define the function $c_r^{\leq i}(x,d)
  = c_r\Big(\sum_{j \in \{1,\dots,i\}:r \in x_j} d_j\Big)$ and rewrite
  $P(x,d)$ as $P(x,d) = \sum_{i \in N} P_i(x,d)$, where $P_i(x,d) =
  U_i(d_i) - \sum_{r \in x_i} c_r^{\leq i}(x,d)$.  Let $G =
  (N,\bar{X},\pi)$ be an arbitrary congestion game with variable
  demands and let $(x,d),(y,\Delta) \in \bar{X}$ be two strategy
  profiles such that $(x,d)=\bigl((x_k,d_k), (x_{-k},d_{-k})\bigr)$
  and $(y,\Delta)=\bigl((y_k,\Delta),(x_{-k},d_{-k})\bigr)$ for some
  $x_k, y_k \in X_k$, $d_k,\Delta_k \in \R_{>0}$, $x_{-k} \in X_{-k}$
  and $d_{-k} \in \R^{n-1}_{>0}$. We notice that $P_i(y,\Delta) =
  P_i(x,d)$ for all $i<k$. Now consider a player $i>k$. When computing
  $P_i(y,\Delta) - P_i(x,d)$, we observe that all costs corresponding
  to resources not contained in $x_k \cup y_k$ cancel out.  For each
  resource $r \in (x_i \cap x_k)\setminus y_k$, we have $- c_r^{\leq
    i}(y,\Delta)+c_r^{\leq i}(x,d) = a_r\,d_k\,d_i$.  Analogously, for
  each resource $r \in (x_i \cap y_k)\setminus x_k$, it holds that
  $-c_r^{\leq i}(y,\Delta) + c_r^{\leq i}(x,d) = -a_r\Delta_k\,d_i$.
  For each resource $r \in x_i \cap x_k \cap y_k$, we have $-c_r^{\leq
    i}(y,\Delta) + c_r^{\leq i}(x,d) = a_r(d_k-\Delta_k)d_i$. Hence,
  for all players $i>k$, we have
\begin{align*}
P_i(y,\Delta) - P_i(x,d)= -\sum_{r \in x_i \cap y_k} a_r\,\Delta_k\,d_i +\sum_{r \in x_i \cap x_k} a_r\,d_k\,d_i.
\end{align*}
 Moreover, we can
calculate straightforwardly that
\begin{align*}
P_k(y,\Delta) - P_k(x,d) &= U_k(\Delta_k) -U_k(d_k) - \Delta_k\sum_{r \in y_k} c_r\Big(\sum_{j \in \{1,\dots,k-1\} : r \in y_j}d_j +\Delta_k\Big)\\
&\quad+ d_k\sum_{r \in x_k} c_r\Big(\sum_{j \in \{1,\dots,k-1\} : r \in x_j}d_j +d_k\Big) - \Delta_k\sum_{r \in y_k} b_r + d_k\sum_{r \in x_k} b_r.
\end{align*}
We thus obtain
{\allowdisplaybreaks
\begin{align*}
  &P(y,\Delta) - P(x,d)\\
  &\quad= \sum_{i \in N} P_i(y,\Delta) - \sum_{i \in N} P_i(x,d)\\
  &\quad= U_k(\Delta_k) -U_k(d_k) -\Delta_k \sum_{r \in y_k} a_r\Big(\sum_{j \in \{1,\dots,k-1\} : f \in x_j}d_j + \Delta_k \Big) + d_k \sum_{r \in x_k} a_r\Big(\sum_{j \in \{1,\dots,k-1\} : r \in y_j}d_j +d_k\Big)\\
  &\quad\quad -\Delta_k\sum_{r \in y_k} b_r + d_k\sum_{r \in x_k} b_r + \sum_{i = k+1}^{n} \left(\sum_{r \in y_i \cap y_k} a_r\,\Delta_k\,d_i + \sum_{r \in x_i \cap x_k} a_r\,d_k\,d_i\right)\\
  &\quad= \sum_{r \in y_k} a_r\Big(\sum_{j \in N\setminus\{k\} : r \in x_j}d_j +\Delta_k\Big)\Delta_k - \sum_{r \in x_k} a_r\Big(\sum_{j \in N\setminus\{k\} : r \in x_j}d_j +d_k \Big)d_k - \Delta_k\sum_{f \in y_k}b_r + d_k\sum_{r \in x_k}b_r\\
  &\quad= {\pi_k}(y,\Delta) - {\pi_k}(y,d).
\end{align*} 
}
Thus, $P$ is an exact potential of $G$.

Since player $i$'s
marginal utility $U_i'(d)$ is concave, there is $D_i \in \R_{>0}$ such that
$U_i'(d) < 2\,a_c\,d + b$ for all $c \in \C$ and $d > D_i$.  By
Lemma~\ref{lemma:equilibrium_demand}, the demands $d_i > D_i$ are not
payoff maximizing and thus strictly dominated. Hence, we can
effectively restrict the set of demands of player $i$ to $[0,D_i]$.
Because the potential function $P$ is continuous, $P$ attains its
maximum on $\varprod_{i \in N} \bigl(X_i \times [0,D_i]\bigr)$ and we
derive that every congestion game with variable demands and affine
costs admits a PNE.

To show the $\alpha$-FIP, let $\bar{x}^{0} \in \bar{X}$ be arbitrary
and let $\gamma = (\bar{x}^{0}, \bar{x}^{1}, \dots)$ be an
$\alpha$-improvement path starting in $\bar{x}^{0}$. For the maximum
of $P$ we have $m = \max_{\bar{x} \in \varprod_{i \in N}
  (X_i,[0,D_i])]} P(\bar{x}) = \max_{\bar{x} \in \varprod_{i \in N}
  (X_i,[0,\R_{\geq 0}])]} P(\bar{x})$. Since $P$ is an exact potential
we have $P(\bar{x}^{k+1}) - P(\bar{x}^{k}) > \alpha$. If $\gamma$ is
infinite, we obtain that $P(\bar{x}^{k}) > m$ for all $k > \bigl(m -
P(\bar{x}^0)\bigr)\, / \, \alpha$, which is a contradiction.
\end{proof}

\paragraph{Homogeneously Exponential Cost Functions}
In the following lemma, we will prove that for every improvement step
at least one of the following actions also yields an improvement:
$(i)$ adapting the configuration only while leaving the demand
constant, $(ii)$ adapting the demand only while leaving the used
resources constant. Formally, we call a subset $I'\subseteq I$ of
improving moves \emph{essential} if $\{\bar{y} : (\bar{x}, \bar{y})
\in I'\} = \emptyset$ implies $\{\bar{y} : (\bar{x}, \bar{y}) \in I\}
= \emptyset$ for all $\bar{x} \in \bar{X}$. Such subsets exist since
the set of improving moves $I$ itself is essential. 

A function $P:\bar{X} \to \R$ is called a 
\emph{local essential potential} of $G$ if
$\bar{x} \in argmax_{\bar{y} \in\bar{X}}P(\bar{y})$
implies $\{\bar{y} : (\bar{x}, \bar{y}) \in I'\} = \emptyset$. Clearly, $\bar{x}$  is a
PNE.
\begin{lemma}
\label{thm:essential_subset}
Let $G$ be congestion game with variable demands and homogeneously
exponential costs. Then, $I' = \bigl\{\bigl((x,d),(y_i,x_{-i},
d)\bigr) \in I\bigr\} \cup \bigl\{\bigl((x,d),(x, \Delta_i,
d_{-i})\bigr) \in I\}$ is an essential subset of improving moves.
\end{lemma}

\begin{proof}
  For contradiction, let $i \in N$ and $(y_i,\Delta_i) \in
  \bar{X}_i$ be such that
  $\bigl((x,d),(y_i,x_{-i},\Delta_i,d_{-i}\bigr) \in I$ but both
  $\bigl((x,d),(y_i,x_{-i},d)\bigr) \not\in I'$ and
  $\bigl((x,d),(x,\Delta_i,d_{-i})\bigr) \not\in I'$. Then, we obtain
\begin{align}
&\pi_i\bigl((x_i,\Delta_i),\bar{x}_{-i}\bigr)\!-\!\pi_i\bigl((x_i,d_i),\bar{x}_{-i}\bigr) = U_i(\Delta_i)\!-\! U_i(d_i)\!-\!\Delta_ie^{\phi \Delta_i}\!\!\sum_{r \in x_i} a_r\,e^{\phi \ell_r(\bar{x}_{-i})}\!+\!d_i e^{\phi d_i}\!\!\sum_{r \in x_i} a_r\,e^{\phi \ell_r(\bar{x}_{-i})}  \leq 0,\label{eq:exp1}\\
&\pi_i\bigl((y_i,d_i),\bar{x}_{-i}\bigr)\!-\!\pi_i\bigl((x_i,d_i),\bar{x}_{-i}\bigr) = \phantom{U_i(\Delta_i)\!-\!U_i(d_i)} - d_ie^{\phi d_i}\!\!\sum_{r \in y_i} a_r\,e^{\phi \ell_r(\bar{x}_{-i})}\!+\!d_ie^{\phi d_i}\!\!\sum_{r \in x_i} a_r\,e^{\phi \ell_r(\bar{x}_{-i})}  \leq 0,\label{eq:exp2}\\
&\pi_i\bigl((y_i,\Delta_i),\bar{x}_{-i}\bigr)\!-\!\pi_i\bigl((x_i,d_i),\bar{x}_{-i}\bigr) = U_i(\Delta_i)\!-\!U_i(d_i)\!- \!\Delta_ie^{\phi \Delta_i}\!\!\sum_{r \in y_i} a_r\,e^{\phi \ell_r(\bar{x}_{-i})}\!+\!d_ie^{\phi d_i}\!\!\sum_{r \in x_i} a_r\,e^{\phi \ell_r(\bar{x}_{-i})}  > 0,\label{eq:exp3}
\end{align}
where the latter inequality expresses the fact that $\bigl((x,d),(y_i,x_{-i}, \Delta_i, d_{-i})\bigr) \in I(i)$. Subtracting \eqref{eq:exp1} from \eqref{eq:exp3} we obtain
$-\Delta_i e^{\phi \Delta_i} \left(\sum_{r \in y_i} a_r\,e^{\phi \ell_r(\bar{x}_{-i})} - \sum_{r \in x_i} a_r\, e^{\phi \ell_r(\bar{x}_{-i})}\right) >0,
$ which is a contradiction to \eqref{eq:exp2}.
\end{proof}

We will use the above Lemma to prove that every congestion game with
variable demands and homogeneously exponential costs admits a local essential potential and thus a PNE.

\begin{proposition}\label{pro:exponential}
  Let $\C$ be a set of homogeneously exponential cost functions. Then,
  $\C$ is consistent.
\end{proposition}

\begin{proof}
  Let $G$ be a game with homogeneously exponential cost functions.
  Consider the function $\Phi(x,d) = \sum_{i \in N} \int_0^{d_i}
  \frac{U_i'(s)}{\phi s+1} ds - \sum_{r \in R} c_r(\ell(x,d))$.  With
  the same arguments as in the proof of Proposition~\ref{pro:potential} we
  can find $D_i \in \R_{>0}$ such that $d_i^{*}< D_i$ holds for all $i
  \in N$in every equilibrium $(x^{*},d^{*})$ of $G$. We claim that
  $\bar{x}^* = (x^*,d^*) \in \arg \max_{(x,d) \in \varprod_{ i \in N}
    (x_i,[0,D_i])} P(x,d)$ is a PNE. Referring to
  Lemma~\ref{thm:essential_subset} it is sufficient to show that there
  is no improving move from $(x^*,d^*)$ in which exclusively either
  the demand or the configuration of a single player is adapted.
 
  Because $(x^*,d^*)$ maximizes $\Phi$, the optimality conditions give
  $\pfrac{\d \Phi(x^*,d^*)}{\d d_i^*} = 0$, if $d_i^* > 0$ and
  $\pfrac{\d \Phi(x^*,d^*)}{\d d_i^*} \leq 0$, if
  $d_i^*=0$. We thus obtain the equations $U_i'(d_i^*) = (\phi d_i^* +
  1)\sum_{r \in x_i} a_r e^{\ell_r(x^*,d^*)} = \pfrac{\d d_i^*\sum_{r
      \in x_i} a_r e^{\phi \ell(x^*,d^*)}}{\d d_i^*}$, if $d_i^* > 0$,
  and $U_i'(d_i^*) \leq (\phi d_i^* + 1)\sum_{r \in x_i} a_r
  e^{\ell_r(x^*,d^*)} = \pfrac{\d d_i^*\sum_{r \in x_i} a_r e^{\phi
      \ell(x^*,d^*)}}{\d d_i^*}$, if $d_i^* = 0$.  Hence, the demand
  $d_i^*$ is optimal for player $i$ for the configuration profile
  $x^*$ implying that there is no improving move in which a player
  solely changes her demand.

Suppose that there is a player~$k$ that changes profitably her
configuration from $x_k^*$ to some $y_k$. Then,
\begin{align*}
\Phi\bigl((y_k,d^*_k),\bar{x}^*_{-k}\bigr) -
\Phi\big((x^*_k,d^*_k),\bar{x}^*_{-k}\big) & = \Big(\frac{1}{e^{\phi
    d_k^*}} - 1\Big)\Big(\sum_{r \in y_k} a_r e^{\phi
  \ell(\bar{x}^*)}\Big) + \Big(1 - \frac{1}{e^{\phi
    d_k^*}}\Big)\Big(\sum_{r \in x^*_k} a_r e^{\phi
  \ell(\bar{x}^*)}\Big) \\
& = \Big(\frac{1}{d_k^*} - \frac{1}{d_k^* e^{\phi d_k^*}}\Big)
\Big(-d_k^* \!\!\sum_{r \in y_k} a_r e^{\phi \ell(\bar{x}^*)} + d_k^*
\!\!\sum_{r \in x_k^*} a_r e^{\phi \ell(\bar{x}^*)}\Big) \\
& = \Big(\frac{1}{d_k^*} - \frac{1}{d_k^* e^{\phi d_k^*}}\Big)\Bigl(\pi_k\bigl((y_k,d_k), \bar{x}_{-k}\bigr) - \pi_k\bigl((x^*_k,d_k),\bar{x}_{-k}\bigr)\Bigr) >0,
\end{align*}
which is a contradiction to the fact that $(x^*,d^*)$ maximizes
$\Phi$. We derive that $(x^*,d^*)$ is a PNE.
\end{proof}

We are now ready to state the main result of this paper.
\begin{theorem}
\label{thm:consistent}
A set $\C \subseteq \twice$ of strictly increasing cost functions is consistent if and
only if one of the following cases holds: $(i)$~$\mathcal{C}$ contains only affine functions $c(\ell) = a\, \ell + b$, $a>0$,$b \geq 0$; $(ii)$~$\mathcal{C}$ contains only homogeneously exponential functions $c(\ell) = a\,e^{\phi \ell}$, $a,\phi > 0$, where $\phi$ is a universal constant of all functions in $\C$.
\end{theorem}
\begin{proof}
Corollary~\ref{cor:hull}  and Proposition~\ref{pro:variable_exp}
imply the "only if" part. Proposition~\ref{pro:potential} and Proposition~\ref{pro:exponential} prove the "if" part.
\end{proof}
The characterization of the set of FIP consistent cost functions
follows from Proposition~\ref{prop:char-alpha-fip}.
\begin{theorem}
\label{thm:fip_consistent}
A set $\C \subseteq \twice$ of cost functions is FIP consistent if and
only if $\mathcal{C}$ contains only affine functions $c(\ell) = a\, \ell + b$ with $a>0$, $b \geq 0$.
\end{theorem}
Our characterizations also hold for network congestion games thus resolving the question raised in.~\cite{OrdaRomSh93}.

\section{Pure Nash Equilibria in the Uniform Cost Model}
\label{sec:uniform}

In the last section, we assumed that the cost function of a resource
defines a \emph{per-unit-price} for every player on that resource.
Thus, the actual cost for a player is the product of the
per-unit-price and the demand of that player. In this section, we will
assume that the cost functions on the resources define \emph{uniform}
costs (or \emph{per-player costs}), that is, every player pays the
same costs regardless of her demand.

Let $\Congmodel$ be a congestion model and let $(U_i)_{i \in N}$ be a
collection of utility functions. We define the corresponding
\emph{uniform congestion game with variable demands} as the strategic
game $G(\congmodel) = (N, \bar{X}, \pi)$, where $\bar{X} = (X,\R_{\geq
  0})$, $ \pi = \varprod_{i \in N} \pi_i$ and $ \pi_i\bigl(x,d\bigr) =
U_i(d_i) - \sum_{r \in x_i} c_r \big(\ell_r\bigl(x,d\bigr) \big), $
and $\ell_r\bigl((x,d)\bigr) = \sum_{j \in N:r \in x_j} d_j$.

Most surprisingly, in the uniform cost model, there is a congestion game with variable demands and linear cost functions without PNE, as illustrated in the following example.

\begin{example}[Uniform cost game without PNE]
\label{ex:affine}
Consider the congestion model $\Congmodel$, where $N = \{1,2\}$ and $R
= \{1,\dots,46\}$.  Each resource has the cost function $c(\ell) =
\ell$.  We define $X= \varprod_{i \in N} X_i$, where
\begin{align*}
X_1 &= \bigl\{\{r_{1}, \dots, r_{22}\}, \{r_{23}, \dots, r_{39}\}\bigr\}\\
X_2 &= \bigl\{\{r_{23},\dots, r_{30}\} \cup \{r_{40}, \dots, r_{46}\}, \{r_{1}, \dots, r_{4}\} \cup \{r_{31}, \dots, r_{39}\}\bigr\}.
\end{align*}
For ease of exposition, we define $R_{1,2} = \{r_{1}, \dots, r_{4}\}$, $R_{1,0} = \{r_{5}, \dots,r_{22}\}$, $R_{2,1} = \{r_{23},\dots,r_{30}\}$, $R_{2,2} = \{r_{31},\dots,r_{39}\}$, $R_{0,1} = \{r_{40},\dots,r_{46}\}$. The strategy space is depicted in Figure~\ref{figure:affine}.

\begin{figure}[ht]
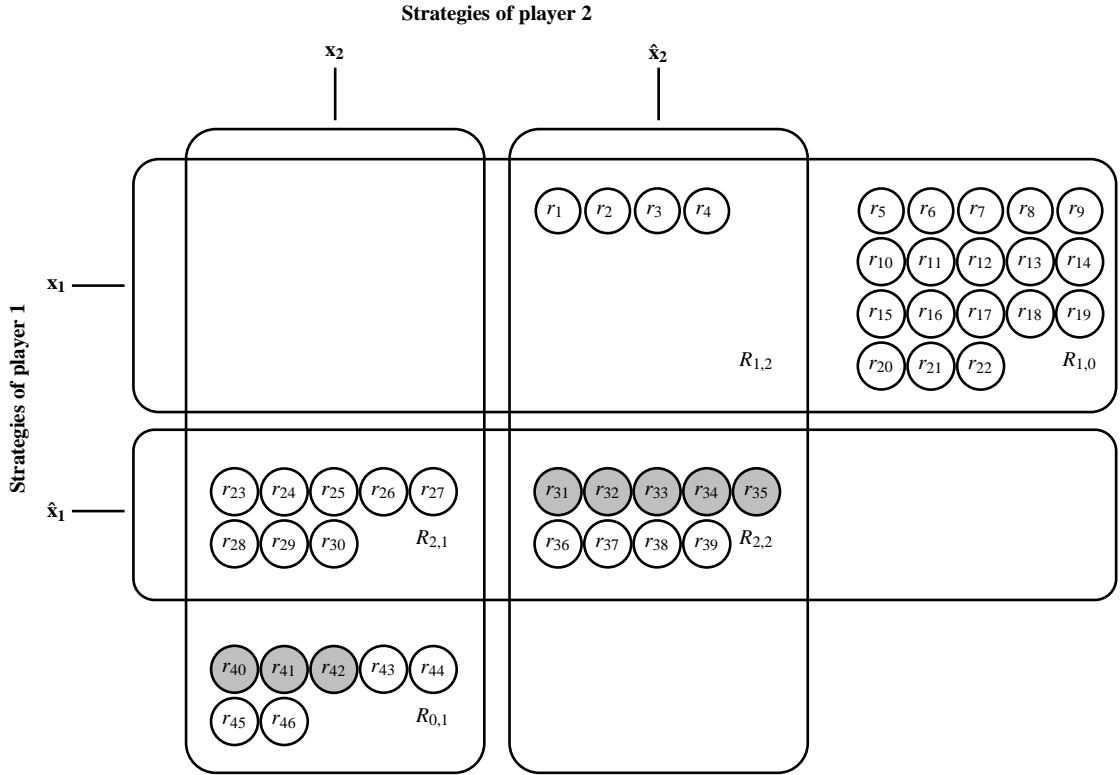

\begin{center}
\psset{arrowsize=5pt, arrowlength=1, linewidth=1pt, nodesep=2pt, shortput=tablr}
\scriptsize
\vspace{1cm}
\hspace{2cm}
\begin{psmatrix}[colsep=0mm, rowsep=0mm, mnode=circle]
%  1 2 3 4 5 - 1 2 3 4 5 - 1 2 3 4 5
  & & & & & &\no\phantom{\hspace{1cm}}&$r_{1~}$&$r_{2~}$&$r_{3~}$&$r_{4~}$& &\no\phantom{\hspace{1cm}}&$r_{5~}$&$r_{6~}$&$r_{7~}$&$r_{8~}$&$r_{9~}$\\
  & & & & & &\no\phantom{\hspace{1cm}}&        &        &        &        & &\no\phantom{\hspace{1cm}}&$r_{10}$&$r_{11}$&$r_{12}$&$r_{13}$&$r_{14}$\\
  & & & & & &\no\phantom{\hspace{1cm}}&        &        &        &        & &\no\phantom{\hspace{1cm}}&$r_{15}$&$r_{16}$&$r_{17}$&$r_{18}$&$r_{19}$\\
  & & & & & &\no\phantom{\hspace{1cm}}&        &        & & &\no$R_{1,2}$   &\no\phantom{\hspace{1cm}}&$r_{20}$&$r_{21}$&$r_{22}$&   &\no $R_{1,0}$\\
\no\\
\no\\
\no\\
\no\\
$r_{23}$&$r_{24}$&$r_{25}$&$r_{26}$&$r_{27}$& &\no\phantom{\hspace{1cm}}&[fillstyle=solid, fillcolor=lightgray]$r_{31}$&[fillstyle=solid, fillcolor=lightgray]$r_{32}$&[fillstyle=solid, fillcolor=lightgray]$r_{33}$&[fillstyle=solid, fillcolor=lightgray]$r_{34}$&[fillstyle=solid, fillcolor=lightgray]$r_{35}$& & & & & &\no\\
$r_{28}$&$r_{29}$&$r_{30}$&  &\no$R_{2,1}$  & &\no\phantom{\hspace{1cm}}&$r_{36}$&$r_{37}$&$r_{38}$&$r_{39}$&\no$R_{2,2}$ & & & & & &\no\\
\no\\
\no\\
\no\\
\no\\
 [fillstyle=solid, fillcolor=lightgray]$r_{40}$&[fillstyle=solid, fillcolor=lightgray]$r_{41}$&[fillstyle=solid, fillcolor=lightgray]$r_{42}$&$r_{43}$&$r_{44}$& &\no\phantom{\hspace{1cm}}& & & & & & & & & & &\no\\
 $r_{45}$&$r_{46}$&  &  &\no$R_{0,1}$  & &\no\phantom{\hspace{1cm}}& & & & & & & & & & &\no\\
\end{psmatrix}
%vertical
\psframe[framearc=0.2](-12.3,-0.7)(-8.3,7.9)
\psframe[framearc=0.2](-8.0,-0.7)(-4.0,7.9)
%horizontal
\psframe[framearc=0.2](-13,7.5)(0.1,4.1)	
\psframe[framearc=0.25](-13,3.9)(0.1,1.6)
\rput[c]{90}(-14.5,4.3){\bf Strategies of player 1}
\rput(-14,5.8){$\bf x_1$}\psline[linewidth=1pt](-13.8,5.8)(-13.1,5.8)
\rput(-14,2.8){$\bf \hat{x}_1$}\psline[linewidth=1pt](-13.8,2.8)(-13.1,2.8)
\rput[c](-8.15,9.4){\bf Strategies of player 2}
\rput(-10.3,8.9){$\bf x_2$}\psline[linewidth=1pt](-10.3,8.7)(-10.3,8.0)
\rput(-6.0,8.9){$\bf \hat{x}_2$}\psline[linewidth=1pt](-6.0,8.7)(-6.0,8.0)
\end{center}
\caption{Strategies of the uniform cost congestion game with variable demands with identity cost and no PNE considered in Example~\ref{ex:affine}. The gray shaded resources are important in the construction of Proposition~\ref{pro:counter_affine}.}
\label{figure:affine}
\end{figure}

For the utility functions $U_1$ and $U_2$ of player 1 and 2,
respectively we chose strictly concave, increasing and twice
continuously differentiable functions satisfying
\begin{align*}
U_1(0)  &= 0,  & U_1(1) &= 32, & U_1(2) &= 53,\\
&& U_1'(1) &= 22, & U_1'(2) &= 17\\[10pt]
U_2(0)  &= 0, &  U_2(1)  &= 32, & U_2(2)  &= 46,\\
&& U_2'(1) &= 15 & U_2'(2) &= 13.
\end{align*}
Clearly, such functions always exists, for instance one can choose
polynomials of degree seven to fit the equations above. These
functions are shown in Figure~\ref{fig:polynomials}.

\begin{figure}[ht]
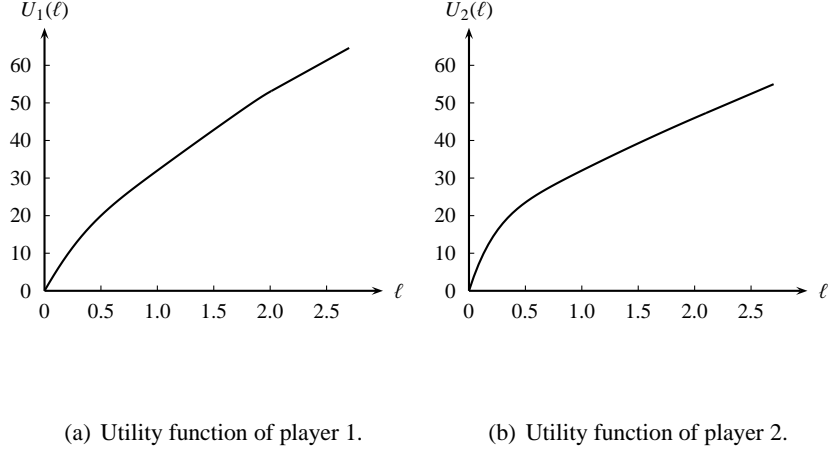

\begin{center}
\scriptsize
\subfigure[Utility function of player $1$.]{%
\psset{xunit=1.5cm,yunit=0.05cm}
\pspicture(0,-5)(3,75)
\psaxes[ticksize=2pt, Dx=0.5,Dy=10]{->}(0,0)(0,0)(3,70)
\psplot{0}{2.01}{
-2.75  x x x x x x x mul mul mul mul mul mul mul
20.5   x x x x x x   mul mul mul mul mul mul
-60.25 x x x x x     mul mul mul mul mul
84.25  x x x x       mul mul mul mul
-46     x x x         mul mul mul
-15.75  x x           mul mul
52     x             mul
add add add add add add 
}
\psplot{2.01}{2.7}{16.62 x mul 19.746 add}
%\psplot{7}{15}{0.5 x -5 add mul 10 add}
%\psplot[linestyle=dotted]{5}{7}{0.5 x -5 add mul 10 add}
%\psplot[linecolor=red]{3}{7}{-0.1875 x mul x mul 3.125 x mul add -1.6875 add}
\rput[l](3.1,0){$\ell$}
\rput[b](0,72){$U_1(\ell)$}
\endpspicture
}
\hspace{1cm}
\subfigure[Utility function of player 2.]{%
\psset{xunit=1.5cm,yunit=0.05cm}
\pspicture(0,-5)(3,75)
\psaxes[ticksize=2pt, Dx=0.5,Dy=10]{->}(0,0)(0,0)(3,70)
\psplot{0}{2.1}{
1       x x x x x x x mul mul mul mul mul mul mul
-10.75  x x x x x x   mul mul mul mul mul mul
49      x x x x x     mul mul mul mul mul
-122.75 x x x x       mul mul mul mul
182.5   x x x         mul mul mul
-162    x x           mul mul
95      x             mul
add add add add add add 
}
\psplot{2.1}{2.7}{12.82 x mul 20.368 add}
\rput[l](3.1,0){$\ell$}
\rput[b](0,72){$U_2(\ell)$}
\endpspicture
}
\end{center}
\caption{Utility functions $U_1$ and $U_2$ of players $1$ and $2$, respectively, considered in Example~\ref{ex:affine}.\label{fig:polynomials}}
\end{figure}

Using that the cost functions are equal to the identity, the
equilibrium conditions for the Nash equilibrium $(x^*,d^*)$ imply
$U_i'(d_i^*) = |x_i|$ for $d_i^*>0$. Said differently, in any PNE in
which player $i$ plays configuration $x_i^*$ and has positive demand
$d_i^*$, the equation $d_i^* = U_i'^{-1}(|x_i^*|)$ holds. Note that
since the utility functions are strictly concave, the derivatives
$U_1'$ and $U_2'$ are convertible and thus their inverse functions
$U_1'^{-1}$ and $U_2'^{-1}$ are well-defined. Now, remark that the
utility functions $U_i$ of player $i$ is chosen such that in
equilibrium $d_i^* \in \{0, 1, 2\}$, since
\begin{align*}
U_1'^{-1}(|R_{1,2} \cup R_{1,0}|)&=1, & U_1'^{-1}(|R_{2,1} \cup R_{2,2}|)&=2,\\
U_2'^{-1}(|R_{2,1} \cup R_{0,1}|)&= 1, &U_2'^{-1}(|R_{1,2} \cup R_{2,2}|)&=2.
\end{align*}
Thus, for any equilibrium $(x^*,d^*)$ we get
\begin{align*}
(x_1^*,d_1^*) &\in \Bigl\{(R_{1,2} \cup R_{1,0},1\bigr), \bigl(R_{2,1} \cup R_{2,2},2\bigr), \bigl(R_{1,2} \cup R_{1,0},0\bigr), \bigl(R_{2,1} \cup R_{2,2},0\bigr)\Bigr\},\\
(x_2^*,d_2^*) &\in \Bigl\{\bigl(R_{2,1} \cup R_{0,1},1\bigr), \bigl(R_{1,2} \cup R_{2,2},2\bigr),\\
&\quad\quad\quad\bigl(R_{2,1} \cup R_{0,1},0\bigr), \bigl(R_{1,2} \cup R_{2,2},0\bigr)\Bigr\}.
\end{align*}
Remark that the latter two strategies of players 1 and 2 with zero demand are indistinguishable in the sense that they always give the same payoff to every player. First, we will show that there is no PNE in which both players chose a strictly positive demand. We calculate that
\begin{align*}
&\pi_1\bigl((R_{2,1} \cup R_{2,2},2),(R_{2,1} \cup R_{0,1},1)\bigr) - \pi_1\bigl((R_{1,2} \cup R_{1,0},1),(R_{2,1} \cup R_{0,1},1)\bigr)\\
&\quad = U_1(2) - 3\cdot|R_{2,1}| - 2\cdot|R_{2,2}| - (U_1(1) -1\cdot|R_{1,2}| - 1\cdot|R_{1,0}|) \\
&\quad = \bigl(53 - 3\cdot 8 - 2\cdot 9\bigr) - \bigl(32 - 1\cdot 4 - 1\cdot 18\bigr) = 11 - 10 = 1,
\intertext{and similarly}
&\pi_2\bigl((R_{2,1} \cup R_{2,2},2),(R_{1,2} \cup R_{2,2},2)\bigr) - \pi_2\bigl((R_{2,1} \cup R_{2,2},2),(R_{2,1} \cup R_{0,1},1)\bigr) = 2 - 1 = 1,\\
&\pi_1\bigl((R_{1,2} \cup R_{1,0},1),(R_{1,2} \cup R_{2,2},2)\bigr) - \pi_1\bigl((R_{2,1} \cup R_{2,2},2),(R_{1,2} \cup R_{2,2}, 2)\bigr) = 2 - 1 = 1,\\
&\pi_2\bigl((R_{1,2} \cup R_{1,0},1),(R_{2,1} \cup R_{0,1},1)\bigr) - \pi_2\bigl((R_{1,2} \cup R_{1,0},1),(R_{1,2} \cup R_{2,2},2)\bigr) = 17 - 16 = 1,
\end{align*}
establishing that neither of these 4 strategy profiles constitutes a PNE. Note that each of these strategy profiles guarantees a strictly positive payoff to each player while choosing $d_i^* = 0$ gives a payoff equal to zero. Hence, both players have an incentive to choose a positive demand which together with the observation that $\gamma$ is an improvement cycle establishes that this game does not admit a PNE.
\end{example}

This is in stark contrast to Proposition~\ref{pro:potential} establishing that every congestion game with variable demand in the proportional cost model possesses a PNE. We will
shed light on this dichotomy by obtaining a complete characterization.
\paragraph{Necessary Conditions.} As in the proportional cost model, we can restrict our search space for consistent cost functions to affine and exponential functions. With a slight modification in the proof of Proposition~\ref{pro:weighted-variable} one can establish that every functions that is not consistent w.r.t. weighted congestion games is also non-consistent w.r.t. uniform congestion games with variable demands. Like this, we get the following immediate corollary.

\begin{corollary}\label{cor:uniform}
 If $\C \subseteq \twice$ is consistent w.r.t. 	uniform congestion games with variable demands then one of the following cases holds: $(i)$ $\C$ contains only affine functions; $(ii)$ $\C$ contains only exponential
  functions.
  \end{corollary}
While the upper result seemingly establishes a structural similarity
between the games with proportional costs, it turns out that they
behave completely different. As already mentioned, uniform congestion games variable demands and affine
costs need not possess a PNE. This holds even for two player games in
which the cost function $c_r$ on all resources $r \in R$ equals the
identity as illustrated in Example~\ref{ex:affine}. Using this example as a blueprint, one can show the following more general result.

\begin{proposition}
\label{pro:counter_affine}
Any affine function is not consistent w.r.t. uniform congestion games with variable demands.
\end{proposition}

\begin{proof}
  In Example~\ref{ex:affine}, we construct a uniform congestion game
  $G$ with variable demands and costs equal to $c(\ell) = \ell$ that
  does not admit a PNE. Now, consider the game $G^{a}$ with utility
  functions $U_1^{a}$ and $U_2^{a}$, where $U_1^{a} = a\,U_1$ and
  $U_2^{a} = a\,U_2$ for some $a>0$. Furthermore, we set the cost on
  all resources equal to $c(\ell) = a\,\ell$. Clearly, the utility of
  every player in $G$ is equal to the scaled utility of that player in
  $G^{a}$. Thus, neither $G$
  nor $G^{a}$ admit a PNE.

  In order to construct a game $G^{a,b}$ with costs equal to $c(\ell)
  = a\,\ell +b$ we introduce a third player with utility equal to
  $U_3^{a,b} = g^{\sigma}\bigl((0,0),\, (b,M)\bigr)$ and a single
  configuration $X_3 = \bigl\{\{31, \dots, 35\} \cup \{40, 41,42\}\bigr\}$.  For
  illustration, these resources are shaded gray in
  Fig~\ref{figure:affine}. With similar arguments as before, we get
  that for sufficiently large $M$ and sufficiently small $\sigma$,
  player $3$ will choose her demand $d_3^*$ in any equilibrium (if
  such exists) such that $d_3^* \in (b-\sigma, b+\sigma)$. We define
  the utility functions $U_1^{a,b}$ and $U_2^{a,b}$ in the game
  $G^{a,b}$ by $U_1^{a,b} = U_1^{a} + 22b$ and $U_2^{a,b} = U_2 +
  18b$. It is easy to check that the difference of the payoffs of the
  games $G^{a,b}$ and $G^{a}$, respectively, goes to $0$ as $\sigma
  \to 0$. Using that the payoff functions of the game $G^{a,b}$ are
  continuous with respect to $\sigma$, we derive that the game
  $G^{a,b}$ does not possess a PNE.
\end{proof}

In order to show a negative result for inhomogenously exponential functions, we will follow the same line of argumentation as for the proportional cost model. To this end, we first introduce the notion of \emph{uniform congestion games with resource dependent demands}. Let $\mathcal{M} = (N, F, X, (c_r)_{r \in R})$ be a congestion model and let $\left(d_{i,r}\right)_{i \in N,r \in R}$ be a matrix of facility-dependent demands. The corresponding \emph{uniform congestion game with resource dependent demands} is the strategic game $G(\mathcal{M})~=~(N, X, \pi)$, where $\pi$ is defined as $\pi=\varprod_{ i \in N} \pi_i$, $\pi_i(x) = \sum_{r \in x_i}  c_r\big(\ell_r(x) \big)$ and $\ell_r(x) = \sum_{j \in N : r \in x_j} d_{j,r}$. 

\begin{proposition}
\label{pro:uniform_rd}
Any inhomogenously exponential function is not consistent w.r.t. uniform congestion games with resource dependent demands.
\end{proposition}

\begin{proof}
Let $c(\ell) = a\,e^{\phi \ell} + b$  with $a, \phi \in \R_{>0}$ and $b \geq -a$ and consider the congestion model $\Congmodel$ with $5$ players and $7$ resources $R = \{r_1,r_2,r_3,r_4,r_5,r_6,r_7\}$. Only the first two players $i \in \{1,2\}$ have two strategies and will be called \emph{non-trivial} players in the sequel of this proof. Their strategy spaces are
given by
\begin{align*}
X_1 &= \{x_1, \hat{x}_1\},& &\text{ where } & x_1 &= \{r_1,r_2\}, & \hat{x}_1 &= \{r_3,r_4\},\\
X_2 &= \{x_2, \hat{x}_2\},& &\text{ where } & x_2 &= \{r_1,r_2,r_3\}, & \hat{x}_2 &= \{r_5,r_6,r_7\}.
\end{align*}
The players $3$, $4$, and $5$ have only one strategy and serve only to increase the costs on some of the resources. They will be called \emph{trivial players}. Their strategies are given by $X_3 = \bigl\{x_3\bigr\} = \bigl\{\{r_1,r_2\}\bigr\}$, $X_4 = \bigl\{x_4\bigr\} = \bigl\{r_4\}$, $X_5 = \bigl\{x_5\bigr\} = \bigl\{\{r_5,r_6,r_7\}\bigr\}$. Observe that the strategies of the two non-trivial players are disjoint. For the non-trivial players $i \in \{1,2\}$ and the parameters $p,q \in \N_{>0}$, we set $d_{i,r} = 2\frac{\ln 2}{\phi}$ for all $r \in x_i$ and $d_{i,r}  = (2 + \frac{p}{q})\frac{ln 2}{\phi}$ for all $r \in \hat{x}_i$.

We set the demand of the non-trivial players $3$, $4$, and $5$ equal to $d_{3,r} = \ln(a_3)\,/\,\phi$, $d_{4,r} = \ln(a_4)\,/\,\phi$, and $d_{5,r} = \ln(a_5)\,/\,\phi$ for all $r \in R$, where $a_3, a_4, a_5 >1$  will be chosen such  that
\begin{multline*}
\gamma = \bigl((x_1,x_2,x_3,x_4,x_5), (\hat{x}_1,x_2,x_3,x_4,x_5), (\hat{x}_1,\hat{x}_2,x_3,x_4,x_5), (x_1,\hat{x}_2,x_3,x_4,x_5), (x_1,x_2,x_3,x_4,x_5)\bigr)
\end{multline*}
is an improvement cycle. Calculating the costs of the respective deviating non-trivial player in $\gamma$, we obtain the following necessary and sufficient conditions
\begin{align}
2^5\,a_3 &> 2^{4+p/q}\,a_4 + 2^{2+p/q} \tag{U1}\label{eq:U1}\\
2^3\,a_3 + 2^{4+p/q}\,a_4 &> 3\cdot2^{2+p/q}\,a_5 \tag{U2}\label{eq:U2}\\
2^{2+p/q}\,a_4 + 2^{2+p/q} &> 2^3\,a_3 \tag{U3}\label{eq:U3}\\
3\cdot 2^{2+p/q}\,a_5 &> 2^5\,a_3 + 2^2\,a_4.\tag{U4} \label{eq:U4}
\end{align}
Combining \eqref{eq:U2} and \eqref{eq:U4} into one inequality, we obtain
\begin{align*}
2\,a_3 + 2^{2+p/q}\,a_4 &> 3\cdot 2^{p/q}\,a_5 > 2^3\,a_3 + a_4\notag\\
\Rightarrow\quad &a_3 < \Biggl(\frac{2}{3}\,2^{p/q} - \frac{1}{6}\Biggr)\,a_4. \tag{U2'} \label{eq:U2prime}
\intertext{Inequalities \eqref{eq:U1} and \eqref{eq:U3} give rise to}
\frac{1}{2}\,2^{p/q}\,a_4 + \frac{1}{8}\,2^{p/q} &<\,a_3 \tag{U1'} \label{eq:U1prime}\\
a_3\,&< \frac{1}{2}\,2^{p/q}\,a_4 + \frac{1}{2}\,2^{p/q} \tag{U3'} \label{eq:U3prime}
\end{align*}
respectively. It is left to show that the inequalities \eqref{eq:U1prime}, \eqref{eq:U2prime}, and \eqref{eq:U3prime} can be satisfied simultaneously. Let us choose $a_4 > \max\{2^{1+p/q}, \frac{3}{2}(1+\frac{1}{2^{p/q}-1})\}$ and $a_3 = \frac{1}{2}\,2^{p/q}\,a_4 + \frac{1}{4}\,2^{p/q}$. Then, \eqref{eq:U1prime} and \eqref{eq:U3prime} are satisfied trivially. As for \eqref{eq:U2prime}, note that
\begin{align*}
\frac{2}{3}\,2^{p/q}\,a_4 - \frac{1}{6}\,a_4 - a_3 = \frac{2}{3}\,2^{p/q}\,a_4 - \frac{1}{6}\,a_4 - \frac{1}{2}\,2^{p/q}\,a_4 - \frac{1}{4}2^{p/q} = \frac{1}{6}\,a_4\,\Biggl(2^{p/q} - 1\Biggr) - \frac{1}{4}2^{p/q} >0
\end{align*}
by the choice of $a_4$. Finally, we set $a_5 = \frac{1}{3\cdot2^{p/q}}\bigl((2^{1+p/q}-1)a_4 -3a_3\bigr)$. It is left to show, that $a_3,a_4$, and $a_5$ are not smaller than $1$. Since $a_4>2^{1+p/q}$ by definition, $a_3 > 2^{p/q} > 1$. As for $a_5$, inequality \eqref{eq:U4} gives rise to $2^{p/q} a_5 > 2\,a_3$ implying $a_5 > 1$.

Thus, we have found $a_3, a_4, a_5>1$ such that $\gamma$ is an improvement cycle, establishing that the corresponding game does not admit a PNE.
\end{proof}

Using the same idea as in the proof of Proposition~\ref{pro:variable_exp} the non-consistency of inhomogenously exponential functions carries over to uniform congestion games with variable demands.

\begin{proposition}
\label{pro:uniform_inhom}
Any inhomogenously exponential function is not consistent w.r.t. uniform congestion games with variable demands.
\end{proposition}

\paragraph{Sufficient Conditions.}

For uniform congestion games with variable demands and homogenous
exponential costs, we establish the existence of a PNE by deriving an
\emph{essential} potential function which generalize the potentials
considered by Monderer and Shapley~\cite{Monderer:1996}. Recall that a
subset $I'\subseteq I$ of improving moves is \emph{essential} if
$\{\bar{y} : (\bar{x}, \bar{y}) \in I'\} = \emptyset$ implies
$\{\bar{y} : (\bar{x}, \bar{y}) \in I\} = \emptyset$ for all $\bar{x}
\in \bar{X}$.  This definition motivates the concept of
\emph{essential generalized ordinal potentials}. Let $G =
(N,\bar{X},\pi)$ be a strategic game. The function $P: \bar{X} \to \R$
is called an \emph{essential generalized ordinal potential} of $G$ if
for all strategy profiles $\bar{x} \in \bar{X}$ there is an essential
subset of improving moves $I'$ such that $(\bar{x}, \bar{y}) \in I'$
implies $P(\bar{y}) > P(\bar{x})$. Clearly, a maximizer of $P$ is a
PNE.

\begin{proposition}
\label{pro:essential_potential_uniform}
Let $G = (N,\bar{X},\pi)$ be a uniform congestion game with variable
demands and homogenous exponential costs. Then, $G$ admits an
essential generalized ordinal potential and possesses a PNE.
\end{proposition}
Combining Corollary~\ref{cor:uniform} and Propositions~\ref{pro:counter_affine},~\ref{pro:uniform_inhom}, and \ref{pro:essential_potential_uniform}, we state the main result of this section.
\begin{theorem}
\label{thm:consistent_uniform}
A set $\mathcal{C} \subseteq \twice$ of cost functions is consistent w.r.t. uniform congestion games with variable demands if and only if $\mathcal{C}$ contains only homogeneously exponential functions. There is no set of FIP consistent functions.
\end{theorem}

%See Proposition~\ref{pro:essential_potential_uniform} for the ``if''-part and Corollary~\ref{cor:uniform}, and Propositions~\ref{pro:counter_affine}~and~\ref{pro:uniform_inhom} for the ``only-if''-part.
%This result follows from Corollary~\ref{cor:uniform}, and Propositions~\ref{pro:counter_affine}~and~\ref{pro:uniform_rd}.

\section{Conclusions and Open Problems}
\label{sec:conclusions}
We considered the fundamental problem of the existence of PNE and convergence properties
of improvement dynamics in
congestion games with variable demands. We obtained several characterizations
of the cost structure with respect to the existence of PNE and the $\alpha$-FIP.
Since our model is general enough to
closely capture many elements of practical applications, we are confident that our results 
help understanding the behavior of myopic play in real systems. 
We conclude the paper by outlining several research directions
that deserve further attention.

While the present work addressed the existence of PNE (and the $\alpha$-FIP)
with respect to the cost structure (without constraining the strategy space 
and the utility functions), it is natural to ask for combinatorial properties 
of the strategy spaces (such as singletons strategies or strategies given by bases of a matroid) 
that ensure the existence of PNE for general cost functions.  Alternatively,
one can restrict the set of feasible utility functions (e.g., assume linear functions)
and ask for the existence of PNE. Also the case of symmetry within the set of players (with respect to their utilities, their strategies or both) is open.

 Another research direction is to investigate the price of anarchy (stability) in congestion
games with variable demands. It would be interesting to characterize the price of anarchy 
for affine cost functions, because in this case there always exists a PNE.

 The design and the analysis of improvement dynamics is fundamental
and deserves further investigation. Because our model captures many elements of
TCP/IP  routing protocols (involving unsplittable routings and elastic demands), it is
of practical importance to further investigate the \emph{stability and scalability} of
distributed improvement dynamics.

\bibliographystyle{plain}
\bibliography{../master-bib}

\end{document}